\documentclass[12pt,preprint]{aastex}

\shorttitle{Binary Stars in 47 Tucanae}
\shortauthors{Albrow et al.}

\begin{document}

\title{The Frequency of Binary Stars in the Core of 47 Tucanae\altaffilmark{1}}

\altaffiltext{1}{Based on observations with the NASA/ESA {\it Hubble Space Telescope} 
                 obtained at ST ScI, which is operated by AURA, Inc. under 
                 NASA contract NAS 5-26555.}

\author{Michael D. Albrow, Ronald L. Gilliland}
\affil{Space Telescope Science Institute}
\affil{3700 San Martin Drive, Baltimore, MD 21218}
\email{albrow@stsci.edu, gillil@stsci.edu}

\author{Timothy M. Brown}
\affil{High Altitude Observatory, National Center for Atmospheric Research}
\affil{P.O. Box 3000, Boulder, CO 80307}
\email{timbrown@hao.ucar.edu}

\author{Peter D. Edmonds}
\affil{Harvard-Smithsonian Center for Astrophysics}
\affil{60 Garden Street, Cambridge, MA 02138.}
\email{pedmonds@cfa.harvard.edu}

\author{Puragra Guhathakurta}
\affil{UCO/Lick Observatory, University of California at Santa Cruz}
\affil{Santa Cruz, CA 95064.}
\email{raja@ucolick.org}
\and
\author{Ata Sarajedini}
\affil{Astronomy Department, Wesleyan University}
\affil{Middletown, CT 06459.}
\email{ata@urania.astro.wesleyan.edu}

\begin{abstract}
Differential time series photometry has been derived for 46422
main-sequence stars in the core of 47 Tucanae. The observations
consisted of near-continuous 160~$s$ exposures alternating between the
F555W and F814W filters for 8.3 days in 1999 July with WFPC2 on the
Hubble Space Telescope.

Using Fourier and other search methods, eleven detached eclipsing binaries and 
fifteen W UMa stars have been discovered, plus an additional ten 
contact or near-contact non-eclipsing systems. After correction for
non-uniform area coverage of the survey, the {\em observed} frequencies of 
detached eclipsing binaries and W UMa's within 90 arcseconds of the 
cluster center are 0.022\% and 0.031\% respectively.

The observed detached eclipsing binary frequency, the
assumptions of a flat binary distribution with $\log$ period and that
the eclipsing binaries with periods longer than about 4 days have
essentially their primordial periods, imply an overall binary
frequency of $13 \pm 6$ \%.  

The observed W UMa frequency and the additional assumptions that W
UMa's have been brought to contact according to tidal circularization
and angular momentum loss theory and that the contact binary lifetime
is $10^{9}$ years, imply an overall binary frequency of
$14 \pm 4$ \%.

An additional 71 variables with periods from 0.4 -- 10 days have been
found which are likely to be BY Draconis stars in binary systems. The
radial distribution of these stars is the same as that of the
eclipsing binaries and W UMa stars and is more centrally concentrated than
average stars, but less so than the blue straggler stars.
A distinct subset of six of these stars fall in an unexpected domain of the
CMD, comprising what we propose to call red stragglers.

\end{abstract}

\keywords{binaries:  eclipsing -- binaries:  general --
globular clusters:  individual (NGC104, 47 Tucanae)}

\section{Introduction}

The central regions of globular clusters represent rich environments
for the study of dynamical processes in stellar systems
\citep{Meylan97}. Of particular interest is the study of the formation
and evolution of close binary stars and their evolutionary products
\citep{Hut92}. These include blue straggler stars
\citep{Shara97}, cataclysmic variables \citep{Edmonds99} , millisecond pulsars 
\citep{Camilo00,Damico01} and low-mass X-ray
binaries \citep{Callanan95}.

Knowledge of the fraction of binary stars in a globular cluster is
critical for the understanding of its evolution. Mass segregation,
caused by two-body relaxation in a cluster, tends to transfer kinetic
energy outwards from the core and move the cluster towards higher
central condensation. This ``gravothermal collapse" can be profoundly
modified by the existence of even a small binary star population
\citep{Heggie92}. Through encounters with single stars, the
gravitational binding energy of binaries can be extracted and
converted into kinetic energy, halting the runaway collapse of the
core. The efficiency of this process depends on the frequency of
binary stars, effectively the ``heat source" for the cluster.

Previous searches for photometric binaries in 47 Tucanae have been
made by \citet{Shara88}, \citet{Edmonds96}, \citet{Kaluzny97a} and
\citet{Kaluzny98a}. Other globular clusters previously surveyed for
binary stars include M71 \citep{Yan94}, M5 \citep{Yan96}, $\omega$ Cen
\citep{Kaluzny96, Kaluzny97b}, M4 \citep{Kaluzny97c}, NGC 4372 
\citep{Kaluzny93} and NGC 6397
\citep{Rubenstein96,Kaluzny97d}.  A summary of contact binaries found
in globular clusters has recently been published by
\citet{Rucinski00}.  All of these surveys suffer from some degree of
limitation.  Rarely have they been deep enough to extend to main
sequence stars, except those right at cluster turnoff. The
ground-based work has been restricted by crowding constraints to
either sparse clusters or regions outside the cores of dense clusters
like 47 Tuc and studies using the Hubble Space Telescope (HST) 
have been limited by observing-time
constraints to relatively short temporal coverage.

Estimates of the binary frequency in globular clusters include the
work of \citet{Yan94} and \citet{Yan96} who derived overall binary
frequencies of 22\% and 28\% (with large uncertainties) for M71 and M5
respectively. These estimates are based on the discovery of a handful
of contact or near-contact binaries, and the assumption of a flat
frequency distribution of binaries with $\log P$. It is also assumed
that initially detached binaries with short periods are brought into
contact through the braking effect of a magnetic stellar wind. In the
dense core of a globular cluster such as 47 Tuc, these assumptions are
subject to modification by the effects of stellar dynamics. Binary
stars, being more massive on average than single stars, will
preferentially sink towards the center of the cluster's gravitational
potential well. In addition, interactions between and among binary and
single stars can result in the formation and destruction of binaries
\citep{Heggie75,Hills75,Hills92,Goodman93,Sigurdsson95,Davies95,Bacon96,
Heggie96,McMillan96,Portegies97}.

Others have taken the route of using single-epoch photometry to derive
binary frequencies from color-magnitude-diagram morphology. This
requires very precise photometry and is also reliant on assumptions
regarding the binary mass-ratio distribution. Estimates of the binary
star frequency by this method include those of \citet{Rubenstein97},
who found likely values in the range 15--38\% for NGC 6752, and
\citet{Elson98} who derived 35\% for the core and 20\% for the outer
parts of the young LMC cluster NGC 1818.

In this paper we present the results of a new search for binary stars
in the core of 47 Tucanae using HST. This is by far the deepest survey
of this type ever made, extending from the main sequence turnoff of
the cluster at $V = 17$, down to magnitudes as faint as $V = 25$.
\S \ref{sectObservations} through \ref{sectOther} will discuss
the observations and results for various classes of binary stars
detected within this primary sample of over 46000 stars that never
saturate the detectors.  For completeness (but not factored into
statistics on frequencies), variables detected using lower quality
photometry on some 3000 saturated stars extending upwards to $V$ = 11
will be discussed in \S \ref{sectGiants}.  Near-continuous
observations for 8.3 days have created the best-sampled time-series
data of any globular cluster observed with HST, or even from the
ground for a comparable temporal window.

\section{\label{sectObservations}Observations}

The data were taken over an 8.3-day period, 1999 July 3-11, during
which HST was pointed continuously at 47 Tuc. The primary goal of the
campaign was a search for close-in, Jupiter-size planets by detection of
transits. The results of this search are described in \citet{Gilliland00}.

Only one field was observed, with the PC1 CCD including cluster
center.  During the observations, a series of 160s exposures were made
with WFPC2, alternating between the F555W and F814W filters. Gaps in
the data occurred during Earth occultations and South-Atlantic-anomaly
passages. A deliberate program of uniform subpixel dithering was
introduced between exposures in order that a spatially-oversampled
mean image could be created from the combined images, and
to provide a stationary distribution of offsets in support of robust
time series. A total of 636
F555W ($V$) and 653 F814W ($I$) exposures were obtained in these 
primary time series.
An additional set of 26 F336W exposures were obtained in orbits with
extended visibility periods (as the target neared the Continuous Viewing
Zone for HST) to provide $U$-band colors and modest time series information.
A few much shorter $V$ and $I$ exposures were also taken to allow 
photometry on stars above the turnoff.

The data frames were initially put through the standard HST
calibration pipeline which involves bias and dark subtraction and
flat-field correction.  For the purposes of creating good star lists
and absolute photometry, image combinations were performed using
methods fully described in Gilliland {\em et al.} (2001, in
preparation).  Briefly, a polynomial representation was developed for
the mean intensity as a function of dither offsets within each pixel
by combining the individual undersampled dithered images and making use of
consistently developed registrations.  An oversampled combined image
was created and the positions and magnitudes of the individual stars
found using DAOPHOT II \citep{Stetson87,Stetson92}. Stars for which
$>$ 90 \% of the light within nominal apertures of radius 5 (PC1), 4
(WF chips) pixels comes from neighbors, or for which the aperture
would include any pixels from bleeding trails on neighbouring
saturated stars, are excluded.  The master star list was generated
based on the F555W and F814W data only.

\begin{figure}
\plotone{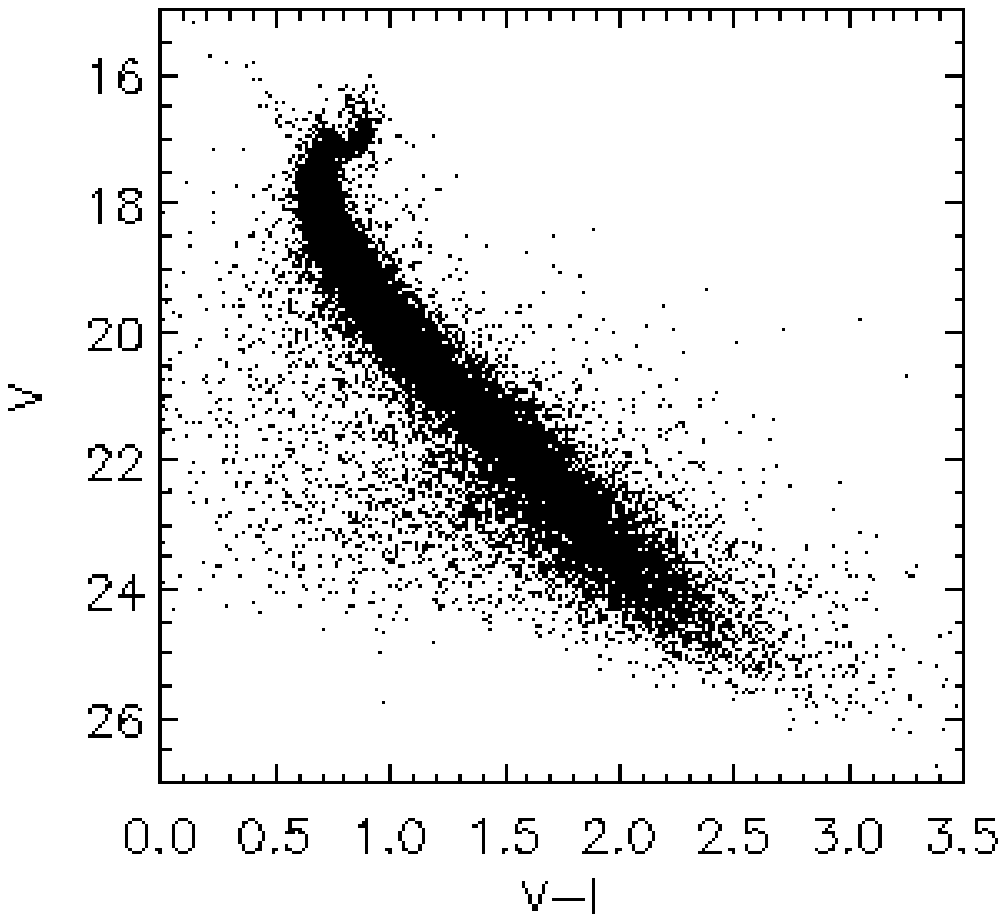}
\caption{\label{CMcombined}
The combined 47 Tuc color-magnitude diagram for all four 
WFPC2 CCDs
for the sample of 46422 stars that remain unsaturated in the
primary exposures.}
\end{figure}

The absolute photometry used here incorporates all standard
corrections for CTE with WFPC2 \citep{Whitmore99} and filter color
terms from \citet{Holtzman95}.  Zero points were determined by
matching the magnitudes of identifiable portions of the cluster main
sequence to those measured for 47 Tuc by other investigators.  $V$ has
been set by forcing $V$ = 17.14 for a first moment centroid
over $0.78 < (V-I) < 0.83$. This value was selected as a
representative average of
\citet{Hesser87} ($V$ = 17.12), unpublished results from RLG
(17.14), \citet{Kaluzny98b} (17.20), and \citet{Alcaino87} (17.22).
For $I$, an adjustment to $V - I$ = 0.69 is forced over 17.48 $<$ $V$
$<$ 17.73, based on values for the position of bluest extent of the
main sequence from \citet{Alcaino87} ($V - I$ = 0.68) and
\citet{Kaluzny98b} (0.71).  Systematic errors of 0.05 magnitudes for
the zero point uncertainty of the absolute photometry scales should be
carried.  The color-magnitude diagram from all four CCDs combined is
shown in Fig.~\ref{CMcombined}.

To generate time series for each star, the oversampled representation
was subtracted from each individual image by evaluating it at the
appropriate dither position and convolving with a compensation kernel,
calculated for each frame to represent focus variations due to thermal
changes in the telescope. Difference-photometry time series were
developed for each star by fitting a PSF (evaluated from isolated
stars in each direct image) at the known star position in the
difference images. The resulting counts in the time series were then
normalised by the total expected in direct images based on
star-by-star magnitudes. These time series were cleaned by removal of
residual, small linear correlations with changes in the
ensemble-average relative intensity and with terms up to cubic in the
combination of \( x \) and \( y \) dither locations.

The difference-image photometry approach (e.g., see \citet{Alard99},
and \citet{Alcock99}), as extended here to under-sampled data,
provides near-optimal results even for stars that are strongly blended
with neighbors.  As shown in \citet{Gilliland00}, time series
precisions range from 0.003 near cluster turnoff at $V \: \sim$
17, to 0.01 at $V \: \sim$ 19.5, 0.03 at $V \: \sim$ 21, and 0.1 at
$V \: \sim$ 23, extending the earlier results beyond the limit for
which giant planets might be detected.  The difference-image
photometry solutions are intimately linked with the process that
develops the over-sampled combined images.  In particular, the
underlying assumption is that the intensity registered for any pixel
can be represented as a function of a constant source and changes only
as a result of $x$, $y$ offsets and focus changes image-to-image.
Thus for the vast majority of non-variable stars precisions near the
fundamental limit imposed by Poisson statistics of the source,
background and readout noise are maintained.  

\begin{figure}
\plotone{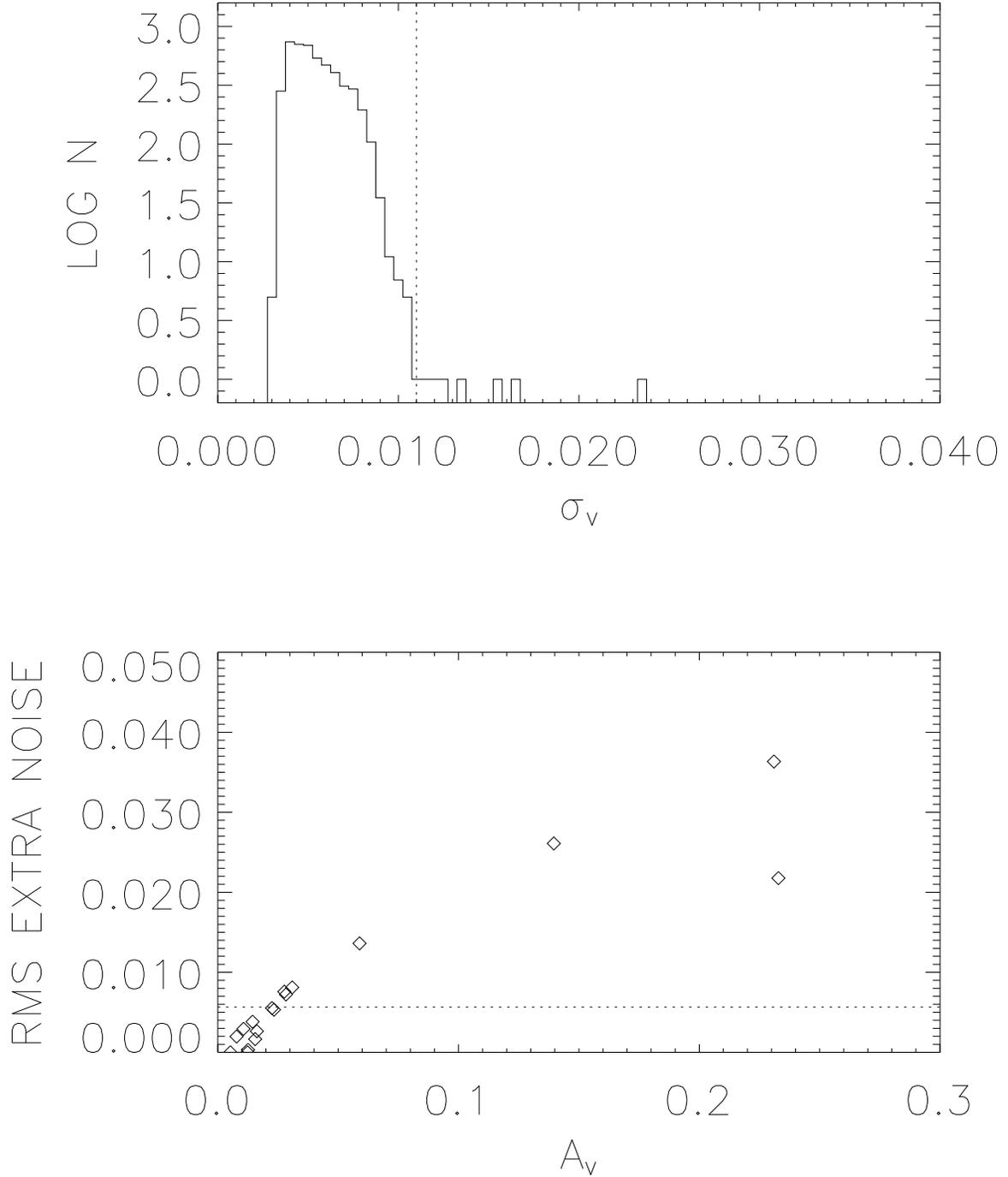}
\caption{\label{ExtraNoise}
Excess ``model'' noise in the photometric time series. Top panel: histogram of 
RMS time series variations for stars with $17 < V < 19$ on PC1. Known
variable stars have been excluded. Lower panel: Excess noise in the 
time series of variable stars in the same magnitude range after subtraction
(in quadrature) of the mean RMS of the non-variable sample. The dotted line
represents the upper envelope of such noise for non-variables.}
\end{figure}

For large-amplitude
variables, excess noise can arise since the solution may not properly
decouple intensity changes in a given pixel that arise from $x$, $y$
offsets of under-sampled PSFs rather than intrinsic variations.  This
makes the time series of some variables noisier than for non-variable
stars at similar magnitudes. We have investigated the characteristics of this 
excess ``model'' noise in some detail. For all the variable stars we detect, we calculate
the magnitude of the model noise by subtracting in quadrature the mean rms for 
non-variables of the same $V$-magnitude from the rms scatter of the variable
star measured from bins of 0.05 in phase. (By measuring in such small phase bins we
eliminate most of the intrinsic variability.)
In Fig.~\ref{ExtraNoise} we plot this excess
noise as a function of amplitude for the variables stars we detected on
the PC1 CCD with $17 < V < 19$. It is clear that the model noise of the detected
variables scales with amplitude of variability. The upper limit
in the non-variable population is 5 mmag. (The few stars that lie above this 
boundary are affected by cosmetics such as diffraction spikes from nearby saturated
stars and show photometric variations at the HST orbital frequency.)
Such plots for other CCDs and 
magnitude ranges show similar behaviour, although the derived upper limit is obviously higher
for fainter magnitudes.

To illustrate the relative unimportance of this excess model noise
for variable star detection, we consider the following scenarios.  At
amplitude 0.1 mag, the model noise is $\sim$ 0.015 mag.  Given our
time coverage with about 600 points in each of the $V$ and $I$
bandpasses, we ask what is the signal-to-noise of a coherent signal
with amplitude 0.1 mag and noise 0.015 mag. From \citet{Scargle82},
the signal-to-noise of the peak periodogram power is $N ( A /
2\sigma)^{2}$, where $N$ is the number of observations, $A$ is the
signal amplitude and $\sigma$ is the noise. If the model noise
dominates the intrinsic rms noise, the S/N in each bandpass is greater
than 6000.  If we consider the regime where $A = 0.02$ mag with model
noise 0.004 mag, a bright star will have a similar level of intrinsic
noise so that the total noise (adding in quadrature) is $\sim$ 0.006
mag. The S/N is then $\sim$ 1000 per bandpass. If the amplitude is
only 0.002 mag, even retaining the same noise level still gives a S/N
of $\sim$ 17 per bandpass.  For fainter stars, the intrinsic noise is
always dominant at low signal levels.  In summary, for small intrinsic
amplitudes, excess model noise is negligible in comparison to the
intrinsic noise; by the point at which large amplitudes may lead to
significant excess noise, the variation is sufficiently large to be
quite obvious.

In many cases the initial variability search returned multiple
variables with similar periods in close spatial proximity.  Where
stars are close enough that their point spread functions overlap to a
significant degree (which is common in such a crowded field) the
extracted time series for each will reflect the variability.  In order
to determine which of several blended stars is the true variable, we
used the measured time variation to choose difference images near both the
maximum and minimum for each variable. We then subtracted the average
of the near-minimum images from the average of the near-maximum
images. This phased, difference-image sum was then compared to a sum
of the direct images for the same exposures.  The true variable
appeared in the difference-based sum as an isolated PSF; comparison
with the direct image then provided a secure identification of the
intrinsic variable.  Instances were found in this way of apparently-bright,
low-amplitude variables having arisen from blending with
nearby fainter variables with intrinsically large amplitudes.  Also,
cases were turned up in which an apparent faint variable resulted from
contamination via the wings of a moderately-distant, saturated,
intrinsically-variable star.

\begin{figure}
\plotone{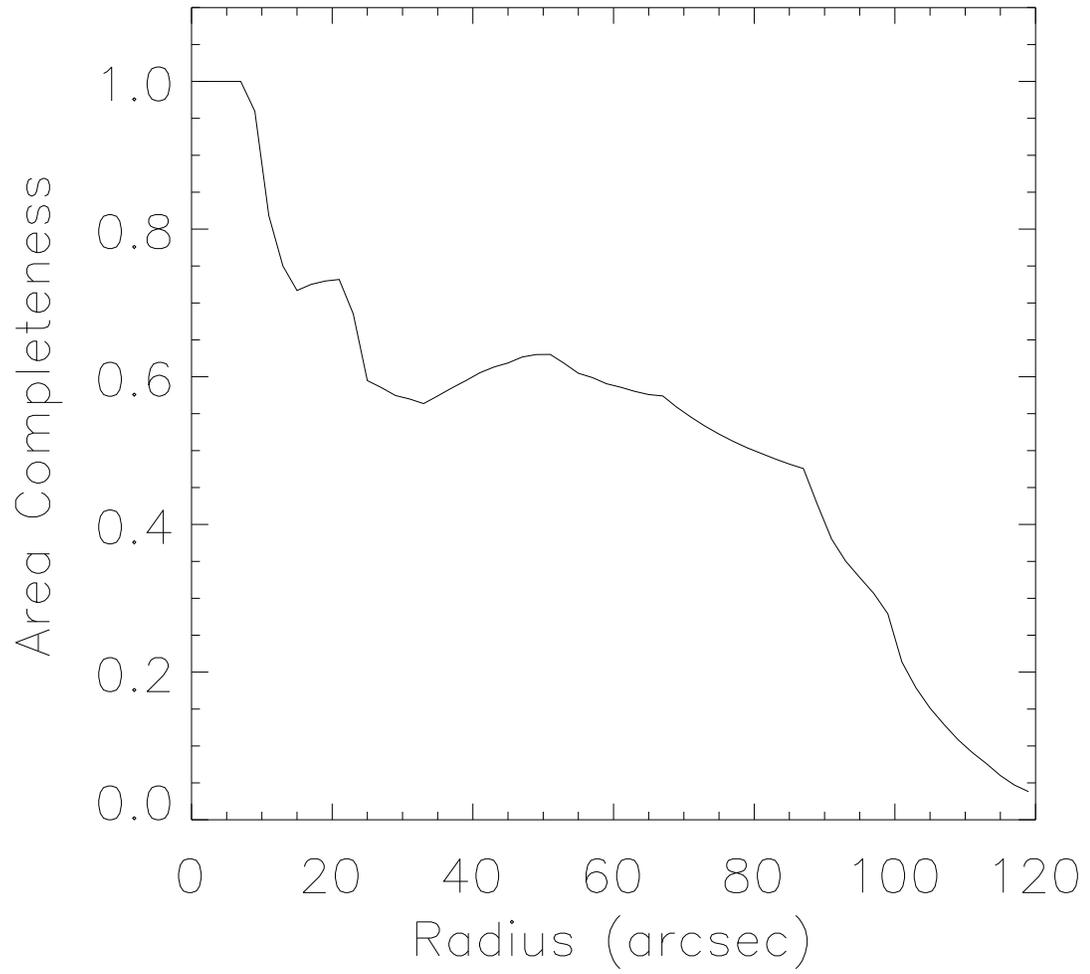}
\caption{\label{RadialArea}
The completeness of the area coverage of our survey as a function 
of radial distance from the center of 47 Tuc.}
\end{figure}

As noted at the beginning of this section, our observations were all taken
at a single pointing, with the center of the globular cluster on the
PC1 CCD. Out to a radial distance of $\sim$ 10 arcseconds, our area coverage is
complete. Beyond this, the non-uniform shape of the array of WFPC2 CCD's meant
that our spatial sampling was lower. In several parts of the remainder of
this paper, we calculate binary-frequency statistics from our observations.
Since we expect binary stars to have a different degree of central
concentration than single stars (this is born out in \S~\ref{sectBYDra}),
we need to be aware of bias which may be introduced by the non-uniform 
radial sampling of our survey. To this end we have calculated the radial
area completeness function (Fig~\ref{RadialArea}) which we use later to correct
our statistics. For consistency with \citet{Howell00}, we take cluster center 
to be at J2000 coordinates (00:24:05.87, -72:04:51.2) from 
\citet{Guhathakurta92}. 
Such a correction implicitly assumes azimuthal symmetry
for the cluster. Normally our correction will be to a circular region
within 90 arcseconds of the cluster center.

\section{Variable Stars}

The time-series data were searched for variability by several
different methods. For near-sinusoidal signals, the Lomb-Scargle (LS)
periodogram is the most useful search technique. For some other types
of variability, such as the lightcurves of detached eclipsing
binaries, the LS periodogram is rather insensitive. In order not to
miss any non-sinusoidal variability we searched for lightcurves where
the mean of the lowest five consecutive points deviated from the
lightcurve mean by a factor of 2 or more times that of the highest
five points (this test is sensitive to eclipsing binaries) and vice
versa (sensitive to microlensing or flaring phenomena).  In practice,
all the variable stars found by this method were also found with the
LS method.  All of the variables found in the primary search (of a
smaller subset of 34091 stars) for planet transits using
matched-filter-convolution approaches that are optimally sensitive to
detecting repeated eclipses (\citet{Gilliland00}, Brown {\em et al.}
(2001), in preparation) were recovered with these independent
searches.

The 160-$s$ time-series exposures were obtained at 240-$s$ intervals,
providing a formal Nyquist sampling frequency of 180~$d^{-1}$.  The LS
periodogram was thus calculated over the period range
\( 0.005-10.0 \) \( d \), frequencies \( 0.1-200 \) \( d^{-1} \).

In order to investigate the appropriate cutoff point for LS false
alarm probabilities in our time series, we have calculated 10 000
Gaussian-noise time series at the same sampling as our data. These
were passed through the same LS analysis and the highest periodogram
peak and its false alarm probability evaluated for each. The lowest
false alarm probability recorded in each of the \( I \)- and
\( V \)- band simulation sets was \( 1.26\times 10^{-4} \). Guided by this
we set the false alarm probability threshold at \( 10^{-4} \) for the
analysis of our data. The lightcurves for all stars in our sample
having a LS false alarm probability less than this were examined in
more detail. We excluded from this procedure periods within \( 1\% \)
of a multiple of \( 1 \) or \( 2 \) times the HST orbital period
(96.66 minutes). In order to pass our test for variability, we required
that a peak greater than the false-alarm probability threshold be present
in either the \( I \)- or \( V \)-band power spectra and that a peak
at the same frequency be recognisable by visual inspection of the
power spectrum of the other band.

\begin{figure}
\plottwo{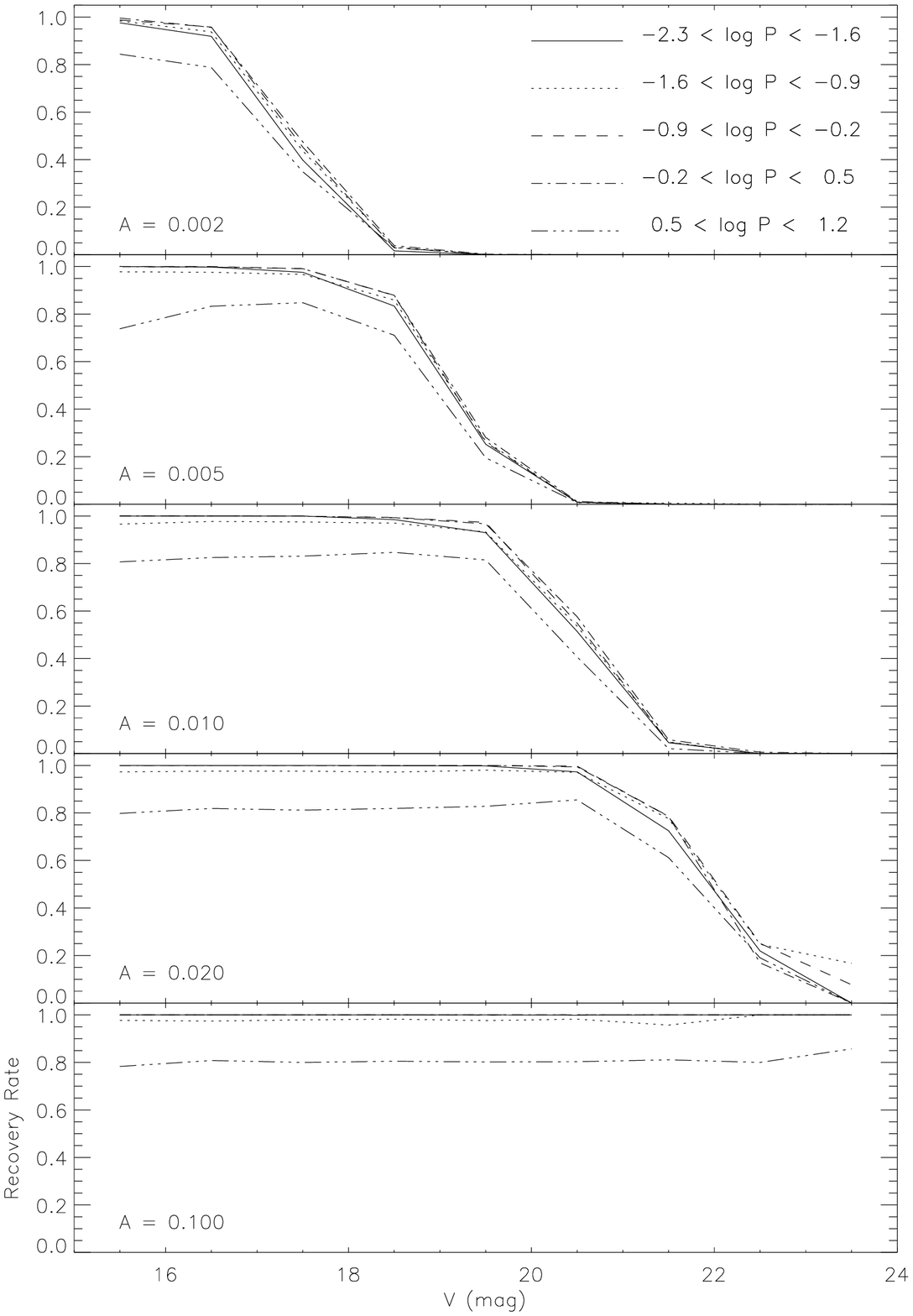}{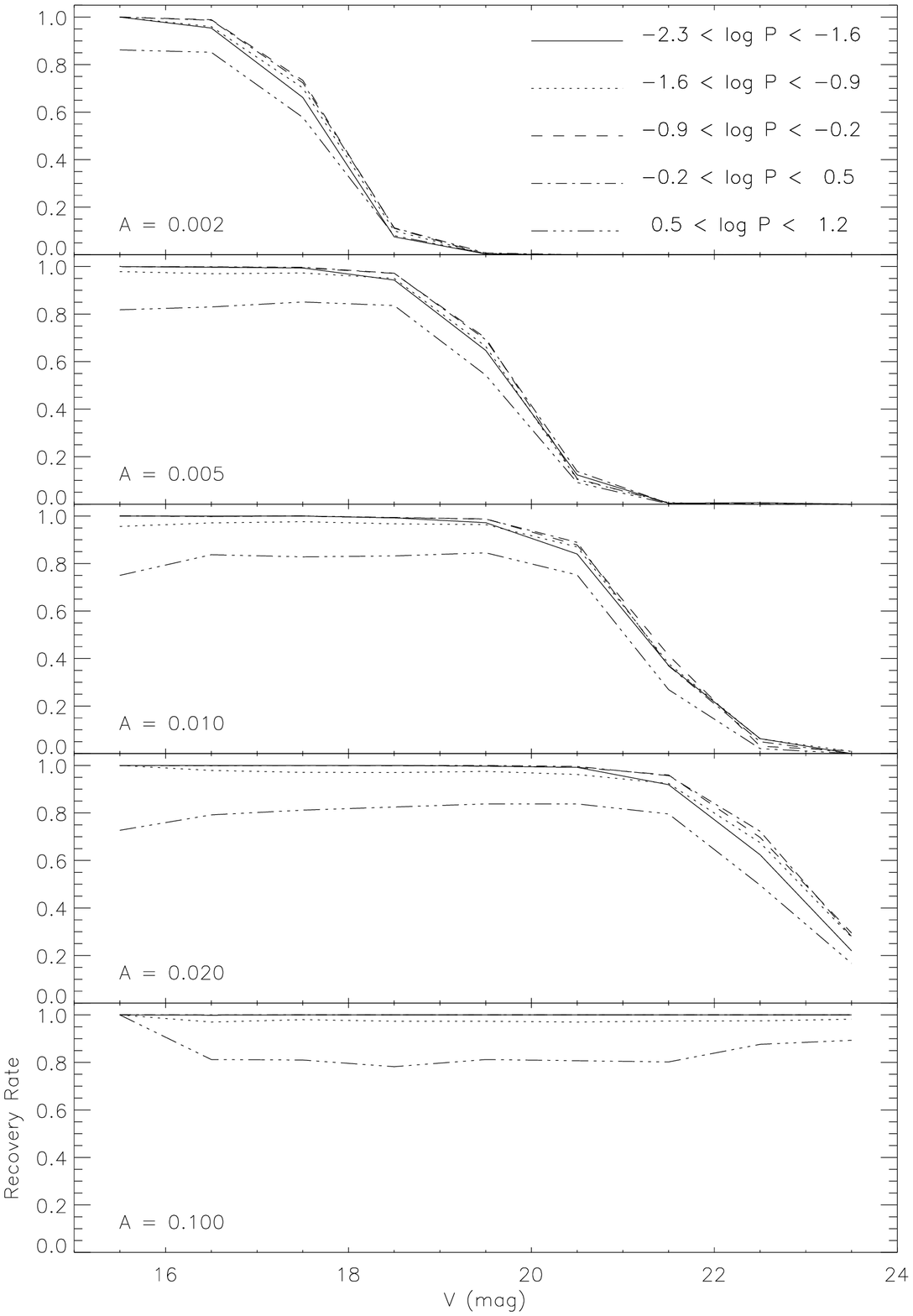}
\caption{\label{RecoveryRate}
Recovery rate of simulated variables as a function of
period, $V$-magnitude and signal amplitude for PC1 (left) and WF2 (right).}
\end{figure}

In order to test our ability to find regular variability, we have
performed tests with simulated data. Sinusoidal signals of random
frequency and phase were injected into the lightcurve data for random
stars (excluding the small fraction of stars which had been found to
be variable) from the PC1 and WF2 CCDs. Because we sample randomly from 
the population of non-variable stars, the actual distribution of noise is 
automatically taken into account.
The chosen periods were drawn
with equal probability in \( \log P \) over the range \( 0.005<P<12 \)
days. The same amplitude was used for \( V \) and \( I \), with
discrete values
\( \frac{\delta (intensity)}{intensity}=(0.002,0.005,0.010,0.020,0.100) \)
and with \( 100,000 \) simulations for each amplitude. The resulting
simulated lightcurves were passed through the same search procedure as
the real data.  From this we have characterized our variability
recovery rate as a function of signal-period, signal-amplitude and
mean \( V \)-magnitude (Fig. \ref{RecoveryRate}).  The recovery rate
is relatively independent of period except at the long period end
where there is a drop-off in sensitivity.

\begin{figure}
\plottwo{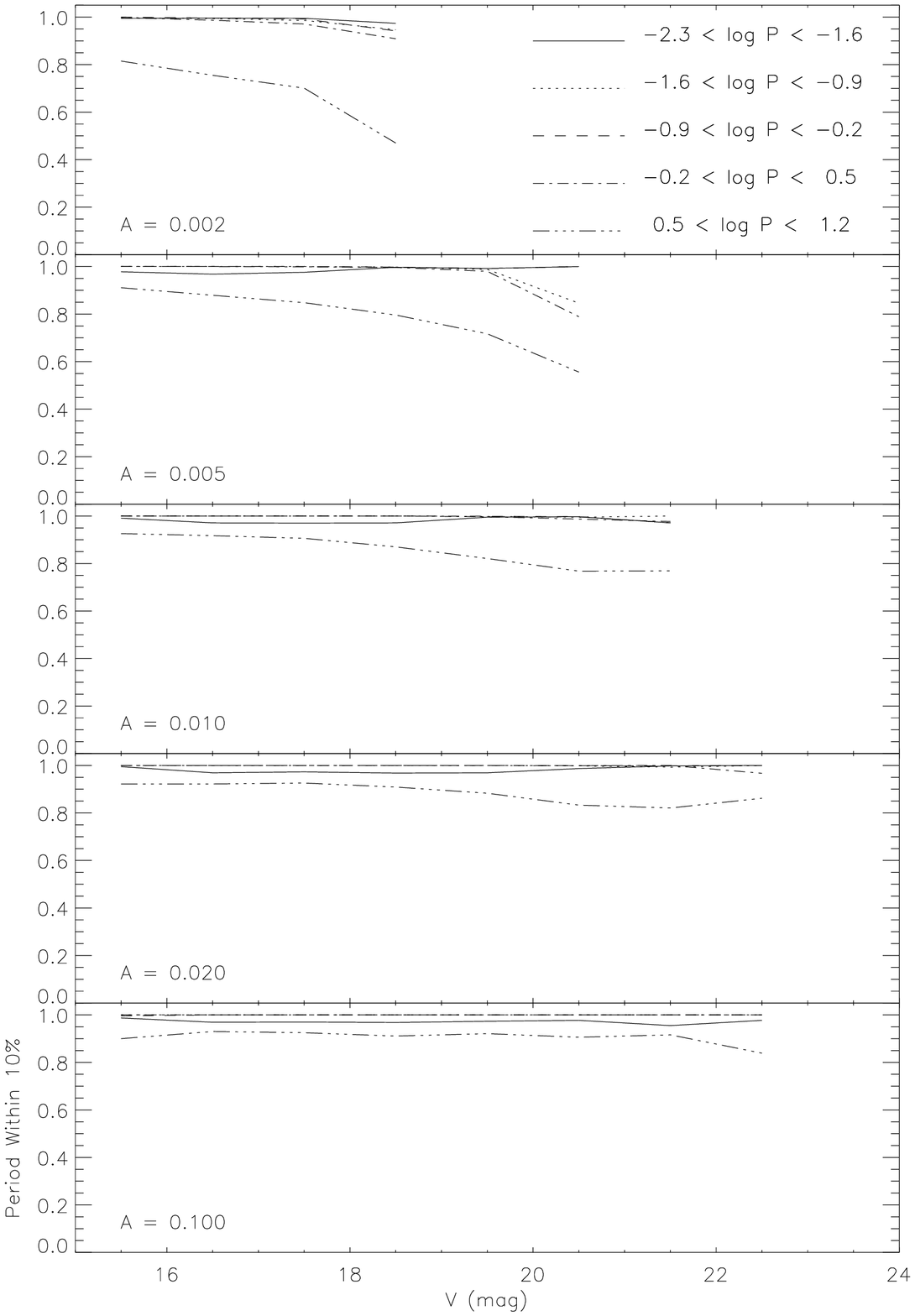}{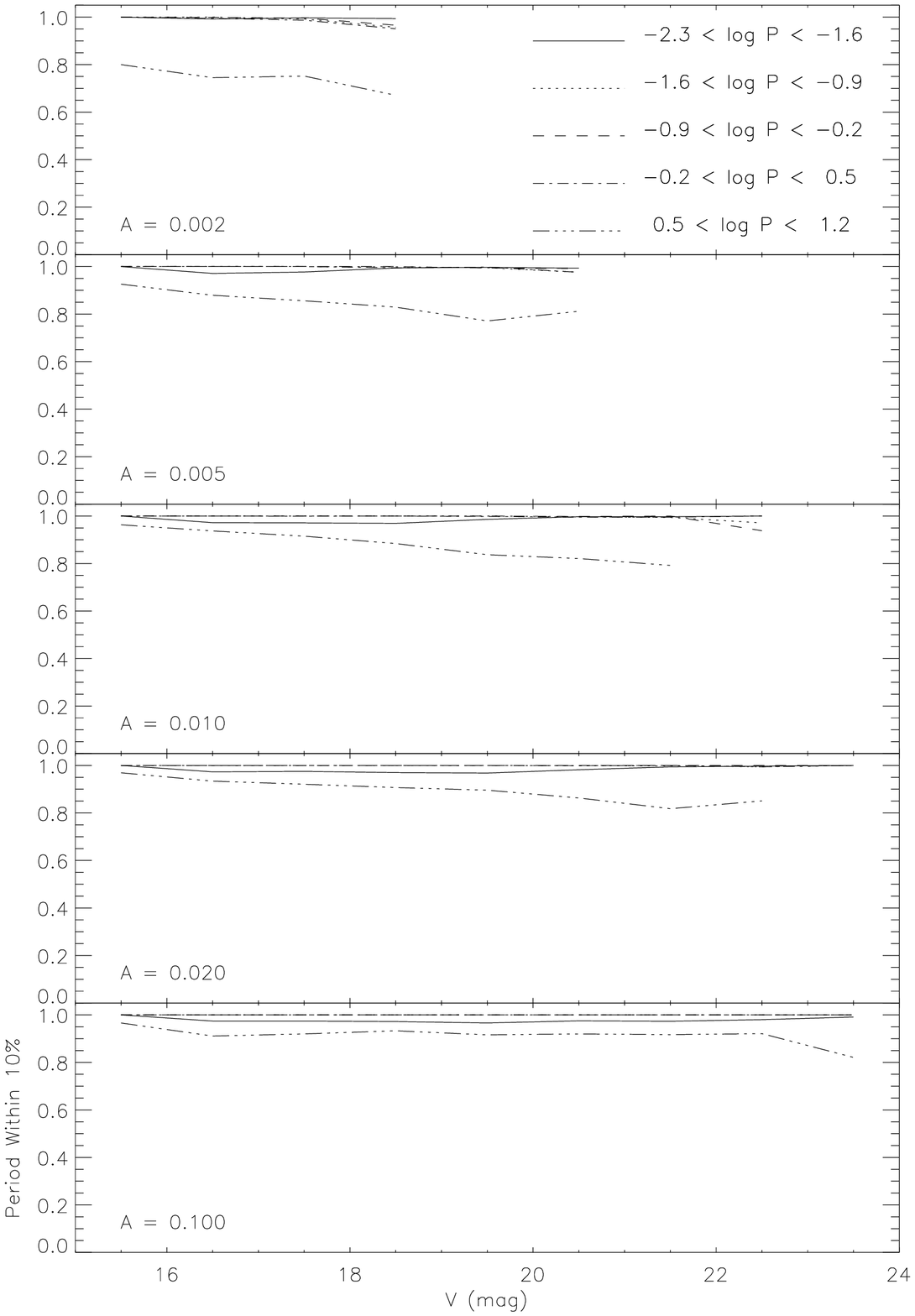}
\caption{\label{TruePeriodRate}
Fraction of recovered variables with periods correct to within
10\% as a function of period, signal-amplitude
and $V$-magnitude for PC1 (left) and WF2 (right).}
\end{figure}

In Fig.\ref{TruePeriodRate} we plot the fraction of the recovered
variables for which the period was determined correctly to within 10\%
of the input period.  There is a rapid decrease at the long-period
end, especially for low-amplitude signals. For example, this implies
that at magnitude \( V=20.5 \) only half of the variables we have
detected with half-amplitudes \( \sim 5 \) mmag will have periods
accurate to within \( 10\% \). This is not as problematical as it may
seem since (see Fig. \ref{RecoveryRate}) our recovery rate is already
low in this regime.

A total of 114 variable stars were detected out of the 46422
non-saturated stars found on the four CCDs.  Nearly all the variable
stars detected are believed to be binaries. At periods from 0.2 days
extending up to 1.1 days we find a number of contact binaries with
obvious W UMa lightcurves. Interspersed in period with these, 
and with periods
extending up to the $\sim$~10 day limit to our sensitivity, are many regular
variables with near-sinusoidal to decidedly non-sinusoidal
lightcurves. We interpret these stars as being BY Draconis variables -
rotating, chromospherically-active main-sequence stars. Since the
timescale for spin-down of a young main-sequence star is vastly
shorter than the age of 47 Tucanae, these stars must also be in
(presumably tidally affected) binary systems. In addition (but not
distinct from the previous category) we find a number of eclipsing
systems.  Finally, there are a few stars which do not fit into these
categories.

In this paper we adopt a convention where we number variable stars
separately for each of the four WFPC2 CCDs. Since the PC1 field
overlaps with our previous observations \citep{Edmonds96,Gilliland98},
we keep the same numbering for the variable stars found in those
studies. Thus (PC1-V01,...,PC1-V16) in this paper are the same stars as
(V1,...,V16) in \citet{Edmonds96} and \citet{Gilliland98}.  For the
other CCDs we begin numbering from unity (e.g. the variables found in
the WF2 field are named WF2-V01, WF2-V02,...).

\section{\label{EBsect} Detached Eclipsing Binaries}

\begin{figure}
\epsscale{0.7}
\plotone{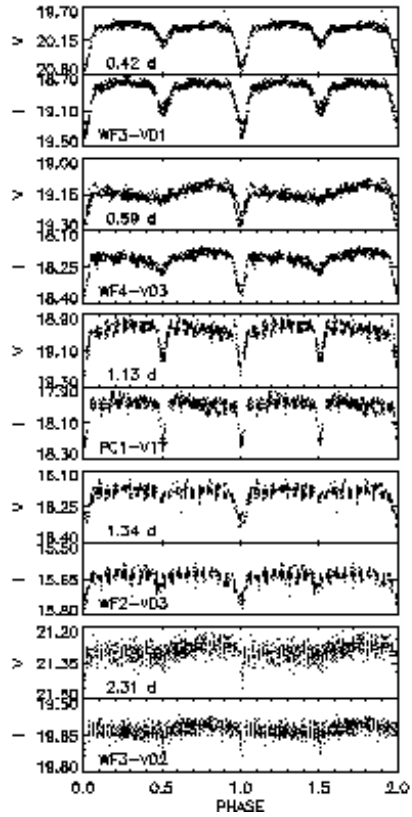}
\caption{\label{lcEB2} 
Lightcurves of eclipsing binaries with periods shorter than 4 days.
Upper panel of pairs provides F555W results with F814W in the lower panel.
Periods in days and reference names corresponding to Table~\ref{tableEB} entries are
displayed for each case.}
\end{figure}

\begin{figure}
\epsscale{0.7}
\plotone{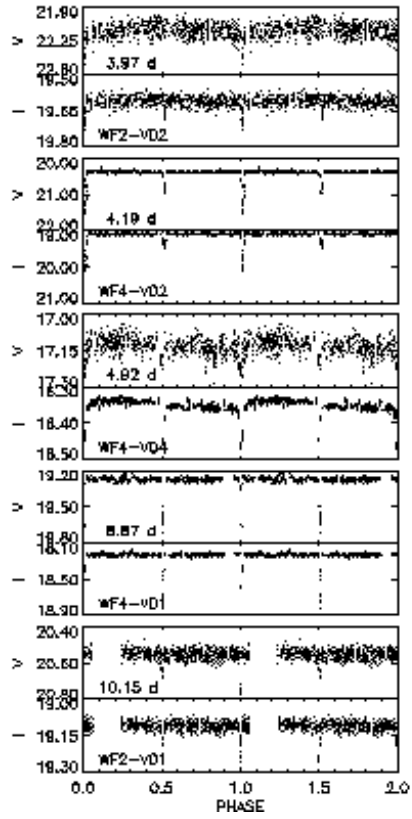}
\caption{\label{lcEB} 
Lightcurves of eclipsing binaries with periods $\gtrsim$ 4 days.}
\end{figure}

Ten of the stars in our sample (Figs~\ref{lcEB2},\ref{lcEB},
Table~\ref{tableEB}), by visual examination, are obvious detached
eclipsing binaries with characteristic primary and secondary dips in
their lightcurves.  An additional eclipsing binary was found which
showed only one eclipse during our observation period and this is
grouped with the miscellaneous variables in \S \ref{sectOther}.
The periods of eclipsing binaries we observe today may not be their
primordial periods. Over time, tidal effects tend to circularize and
then bring together initially-detached binaries. The upper cutoff
period for tidal circularization increases with age and is likely to
be in the range 13--18 days for a population with age of 47 Tucanae
\citep{Mathieu94}.  With subsequent angular momentum loss, many
binaries with periods less than some cutoff period will have evolved
to significantly shorter periods during the lifetime of the cluster.
We have used the model of
\citet{Vilhu82} (see also \citet{Bradstreet94}), through which the 
decrease of spin angular momentum by magnetic braking is coupled to an
increase of orbital angular momentum (and hence a shorter period) to
calculate this cutoff period.  Applied to a 0.9 $M_{\odot}$ star, we
calculate that 4 days is the cutoff period for binaries in 47 Tuc with
mass ratio $q \sim 0.3$. Those with smaller q would have evolved to
contact in less than 11 Gy. (Note that this model does not apply to
very low values of $q$, such as for planets, where there is no
significant tidal affect on the more massive component.)  For binaries
where $q \sim 0.9$, none would have evolved to contact over the
cluster lifetime, assuming they had primordial periods greater than
2.5 days.

We thus consider here the five eclipsing binaries with periods
$\gtrsim$ 4 days as being an ``unevolved sample'' and ask what their
numbers imply about the fraction of primordial binary stars in the
core of 47 Tuc.

We assume that we have detected all the binary stars in our sample
which show two eclipses during the time span of the observations. This
places an upper limit of $\sim$~16.6 days to the periods we could have
detected, assuming that both primary and secondary eclipses are
visible.  Using Kepler's third law and geometry, the probability of
detecting eclipses in a given binary with magnitude $V$ and period $P$
in days is
\begin{equation}
D(V,P) = \frac{\zeta R_{1} + R_{2}}
         {4.21 (M_{1}+M_{2})^{1/3} P^{2/3}} \times
         \max(0,\min(1,\frac{2\tau_{obs}}{P}-1))
\end{equation}
where $M_{1}, M_{2}$ and $R_{1}, R_{2}$ are the mass and radius in
solar units of each component (inferred from $V$), $1 - \zeta$ is the
minimum fraction of the eclipsed star's radius we require to be
obscured for a detectable eclipse and $\tau_{obs}$ is the observation
period, 8.29 days. For any assumed mass ratio, the masses and radii of
the components can be drawn from a suitable isochrone after allowing
for partition of the observed luminosity between each star in the
appropriate ratio.

For a flat primordial frequency distribution of binaries,
$\frac{df}{d \log P} = \beta_{0}$, where $f$ is the cumulative
fraction of binaries with periods less than $P$,
\begin{equation}
f_{0} = \beta_{0} ( \log P_{max} - \log P_{min} ),
\end{equation}
where $f_{0}$ is the fraction of primordial ``hard'' binaries (those
where the orbital velocity is greater than the velocity dispersion of
the cluster). ``Soft'' binaries are expected to be quickly disrupted
through interactions with other stars \citep{Hills84}. The central
velocity dispersion of 47 Tuc is 11.6 km s$^{-1}$ \citep{Meylan86},
implying a maximum orbital period for ``hard'' binaries of $\sim$ 50
years.  (This limit will be smaller if a sufficient binary population
exists to allow the larger cross sections of binary-binary
interactions to be important in disrupting systems.)  
$P_{min}$ = 2.5 d is the minimum observed
period in zero-age binary star populations \citep{Mathieu89}.  Thus,
for an observational sample with $N_{V} dV$ stars in each
$V$-magnitude interval $dV$, the number of eclipsing binaries we
expect to detect over a period range $P_{1}$ to $P_{2}$ and magnitude
range $V_{1}$ to $V_{2}$ is
\begin{equation}
N_{P_{1},P_{2}}^{eclipsing} = \int_{V_{1}}^{V_{2}} 
     \int_{\log P_{1}}^{\log P_{2}} 
     N_{V} D(V,P) \beta_{0} d \log P dV,
\end{equation}
which is a linear function of  $f_{0}$.

\begin{figure}
\epsscale{1.0}
\plotone{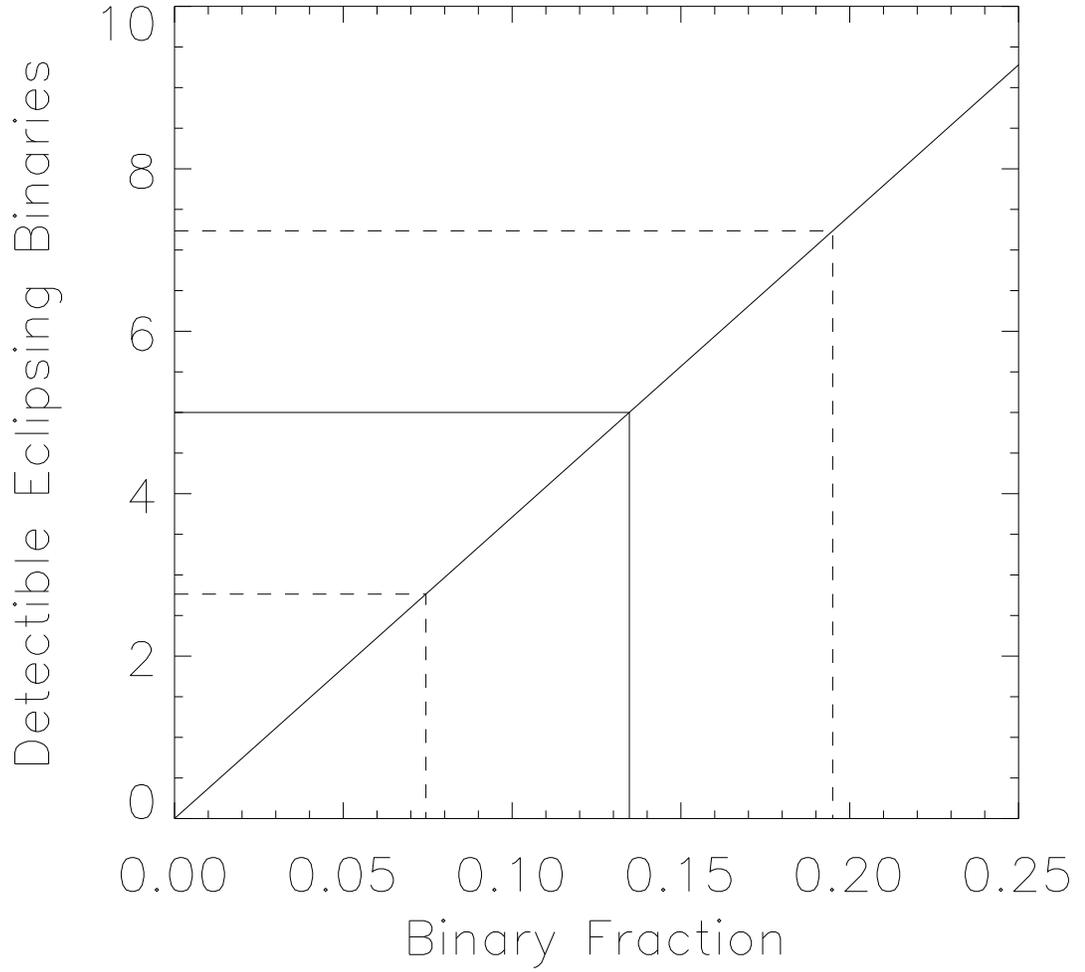}
\caption{\label{eclipse_probability} 
Number of detectable eclipsing binaries vs binary fraction of the
cluster. Indicated is the observed number of eclipsing binaries and
its statistical error.}
\end{figure}

Figure~\ref{eclipse_probability} shows this relation for 47 Tuc, where
we have used the 11 Gy \citet{Vandenberg00} isochrone for the masses
and radii, have assumed equal masses, and that $\zeta = 0.9$. We have
detected five eclipsing binaries with periods between 4 and 16 days,
which implies that the primordial hard binary fraction of the cluster
core is 13 $\pm$ 6 \%. For this calculation and in the remainder of
this paper we define the binary fraction as being the number of binary
stars relative to the number of main sequence stars searched.  The
error we quote here is due entirely to the $\sqrt{n}$ Poisson uncertainty
in
the number of detections, which far outweighs internal errors in any of
the measurements or parameters. For instance, using $\zeta = 0.8$
increases the derived binary fraction by less than 1\% and using a
mass ratio $q = 0.5$ increases the derived binary fraction by
$\sim$2\%.

We can also examine the sensitivity of our model to assumptions about
the binary period cutoff and the shape of the frequency distribution.
If the efficiency of disruption of binaries is much greater than
expected and the upper limit to binary periods is 5 years rather than
50 years, then the binary fraction we derive is 10 $\pm$ 4 \%.  If
the frequency distribution of binaries is not flat, but rather follows
the exponential function of \citet{Duquennoy91} (found from nearby G
dwarfs), $f = \beta_{0} e^{-0.095(\log P -4.8)^2}$, the derived binary
fraction becomes 17 $\pm$ 8 \%.

As noted in \S~\ref{sectObservations}, our observations are subject to
bias because of radial incompleteness in the area coverage. For the
eclipsing binaries considered here, the correction factor is negligible,
changing the estimated binary fraction by < 0.2 \%. 

\begin{figure}
\plotone{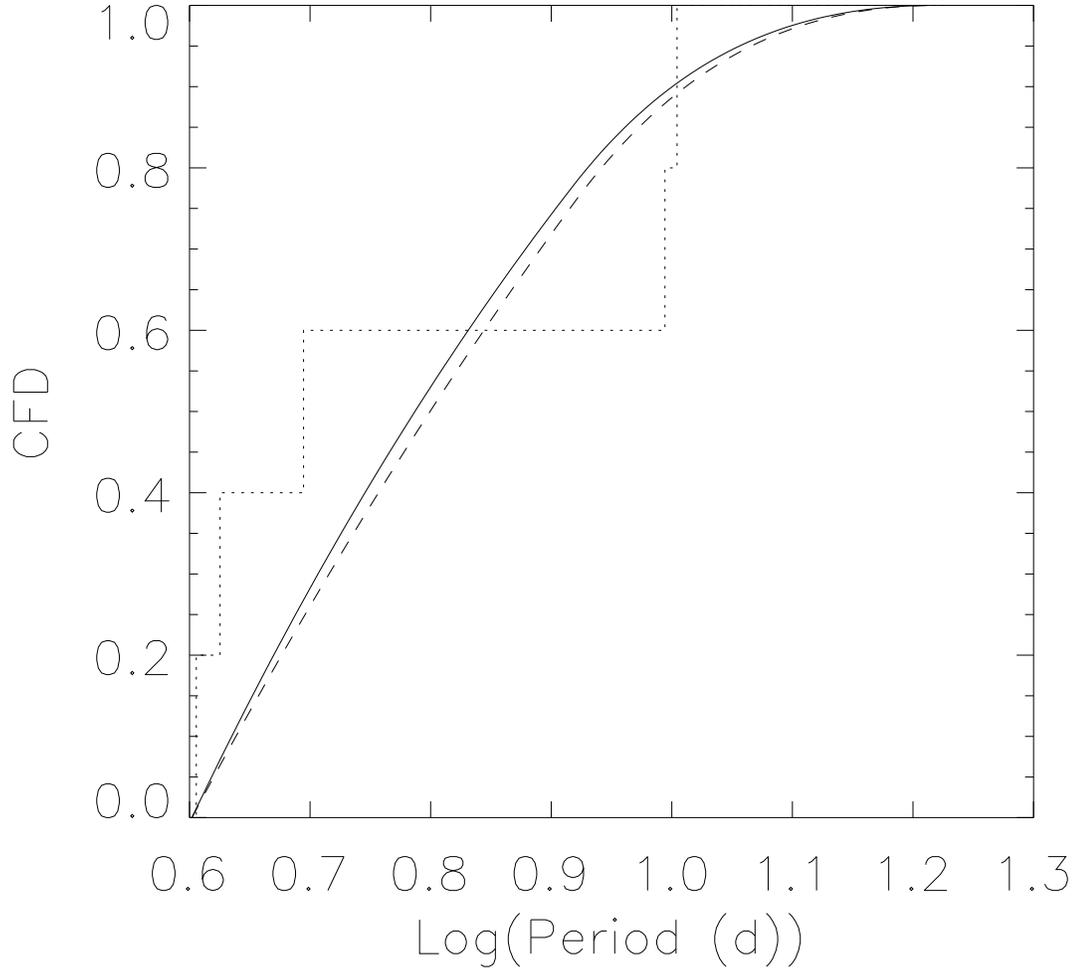}
\caption{\label{eclipsingPfreq} 
Cumulative distributions with period of the expected and observed
eclipsing binaries. The solid curved line is the flat model and the
dashed line the exponential model of \citet{Duquennoy91}.  The dotted
stepped line represents the observed periods.}
\end{figure}

We can ask whether the periods of the stars we have detected are
consistent with flat or exponential distributions. In practice,
$D(V,P)$ is almost independent of V, thus simplifying a search for
period dependence.  Unfortunately two of the five eclipsing binaries
being considered each show only two eclipses during the time of
observation so there is a slight ambiguity as to whether these are
primary and secondary eclipses (in which case the periods are 8.87 and
10.15 d) or primary eclipses only (in which case the periods are half
these values). In fact, detailed inspection of the lightcurves argues
for the former. For WF4-V01, the depths of the two observed $V$--band
eclipses differ. For WF2-V01, the depths of the two observed eclipses
are very similar but there is no apparent eclipse at the phase midway
between them. If there was an unobserved secondary eclipse at this
phase then this would place such a small upper limit on the radius of
the secondary that the primary eclipse would have a flat bottom, which
is not the case.  In Fig.~\ref{eclipsingPfreq} we show the cumulative
period distribution compared with the models for flat and exponential
distributions.  The cumulative distribution is driven mainly by the
geometric factor for probability of eclipses given random
orientations, so in practice does not provide a good discriminant
between models.  The Kolmogorov-Smirnov statistic \citep{Press92} for
the maximum difference between the distributions of the flat model and
observed periods is 0.33 with a formal probability of 54\% that the
observations are drawn from the model population.

\section{\label{WUMasect} W Ursa Majoris Binaries}

\begin{figure}
\epsscale{0.7}
\plotone{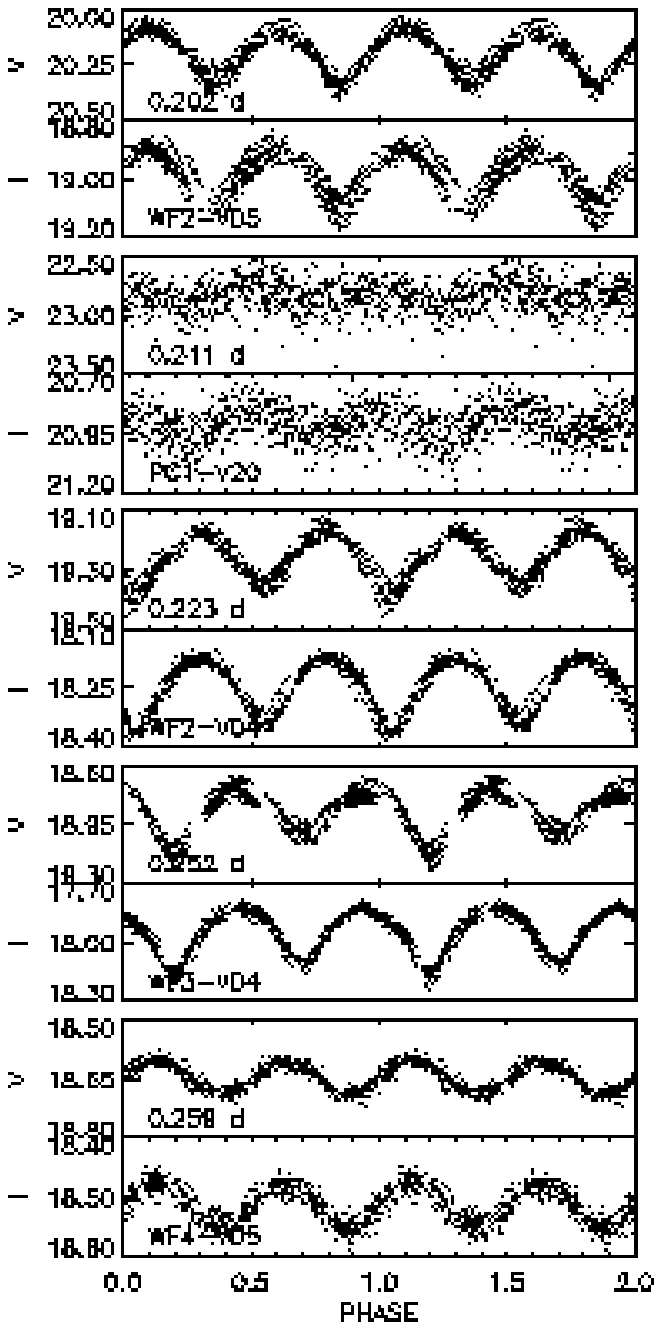}
\caption{\label{lcWUMa1} 
Lightcurves of W UMa and near-contact binaries.}
\end{figure}

\begin{figure}
\epsscale{0.7}
\plotone{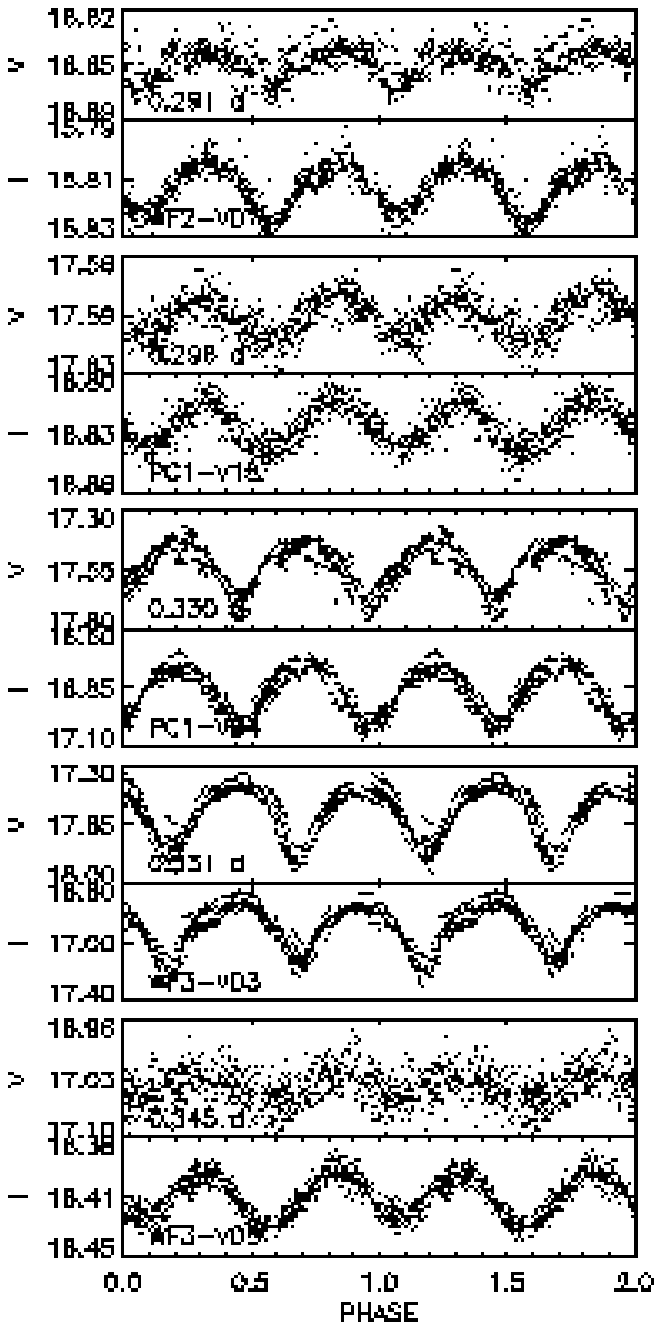}
\caption{\label{lcWUMa2} 
Lightcurves of W UMa and near-contact binaries.}
\end{figure}

\begin{figure}
\epsscale{0.7}
\plotone{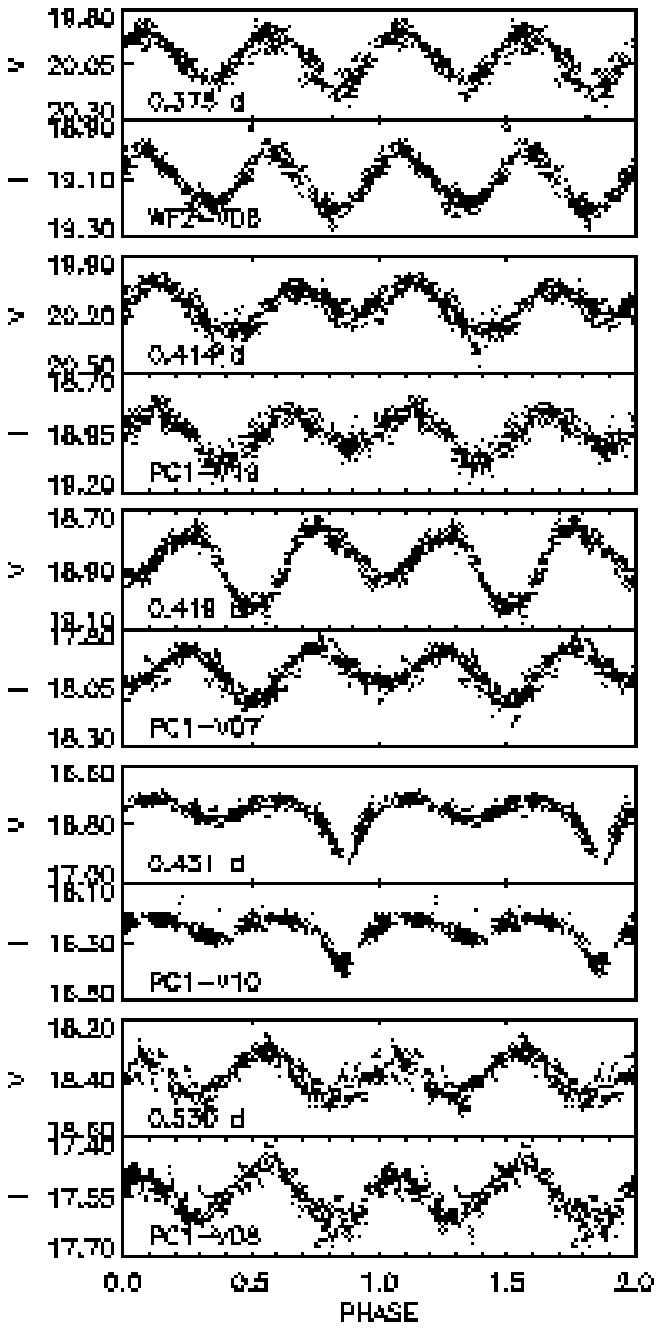}
\caption{\label{lcWUMa3} 
Lightcurves of W UMa and near-contact binaries.}
\end{figure}

Fifteen of the stars in our sample have, by visual examination, characteristic 
W UMa lightcurves (Figs.~\ref{lcWUMa1},
\ref{lcWUMa2}, \ref{lcWUMa3}, Table~\ref{tableWUMa}). These were initially 
assumed to be contact binary systems.  The generally accepted theory
for the formation of such systems is that detached binaries 
lose orbital angular momentum by a magnetic
braking mechanism when their spin and orbital angular momentum become
tidally coupled \citep{Vilhu82, Eggen89}.  The rate of approach
increases as the separation decreases because the strength of the
magnetic stellar wind increases with higher rotational velocities
\citep{Skumanich72}.  For nearly equal mass stars, a rapid initial
mass transfer will occur after first contact is established 
\citep{Flannery76,Robertson77,Iben93} because,
for a main sequence star with a convective envelope, the Roche-lobe
radius shrinks faster in response to mass loss than does the stellar
radius. Subsequent mass-transfer oscillations (for part of which time
the system may be in a semi-detached state) result in a system with
mass ratio varying over the range $0.61 < q < 0.79$.  An important
consequence is that the total luminosity of a nearly-equal-mass
pre-contact binary will increase as the system adjusts to a different
mass ratio \citep{Vilhu82}.  For instance, a binary with initial
component masses of 0.75~$M_{\odot}$ will be brighter by 0.34 mag
after it evolves to a mass ratio $q = 0.8$. The brighter component
will also be bluer leading to a tendency for bright W UMa systems to
be blue stragglers.

\begin{figure}
\epsscale{1.0}
\plotone{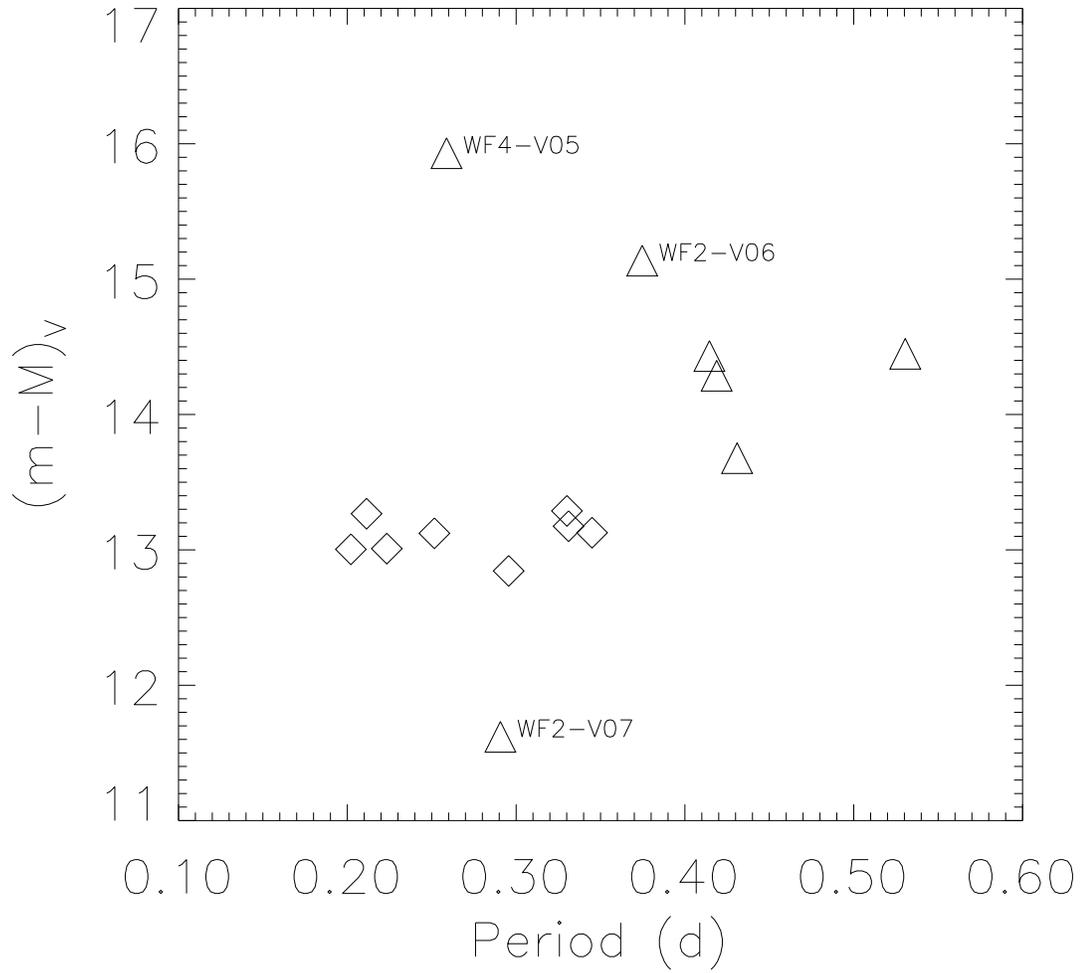}
\caption{\label{DMWUMa} 
Distance modulus of the 14 W UMa and near-contact binaries according to the calibration of
\citet{Rucinski95}.
The eight stars consistent with being in full contact are shown with
diamonds (Group 1 in text), triangles (Group 2) indicate likely 
semi-detached systems.}
\end{figure}

\begin{figure}
\epsscale{1.0}
\plotone{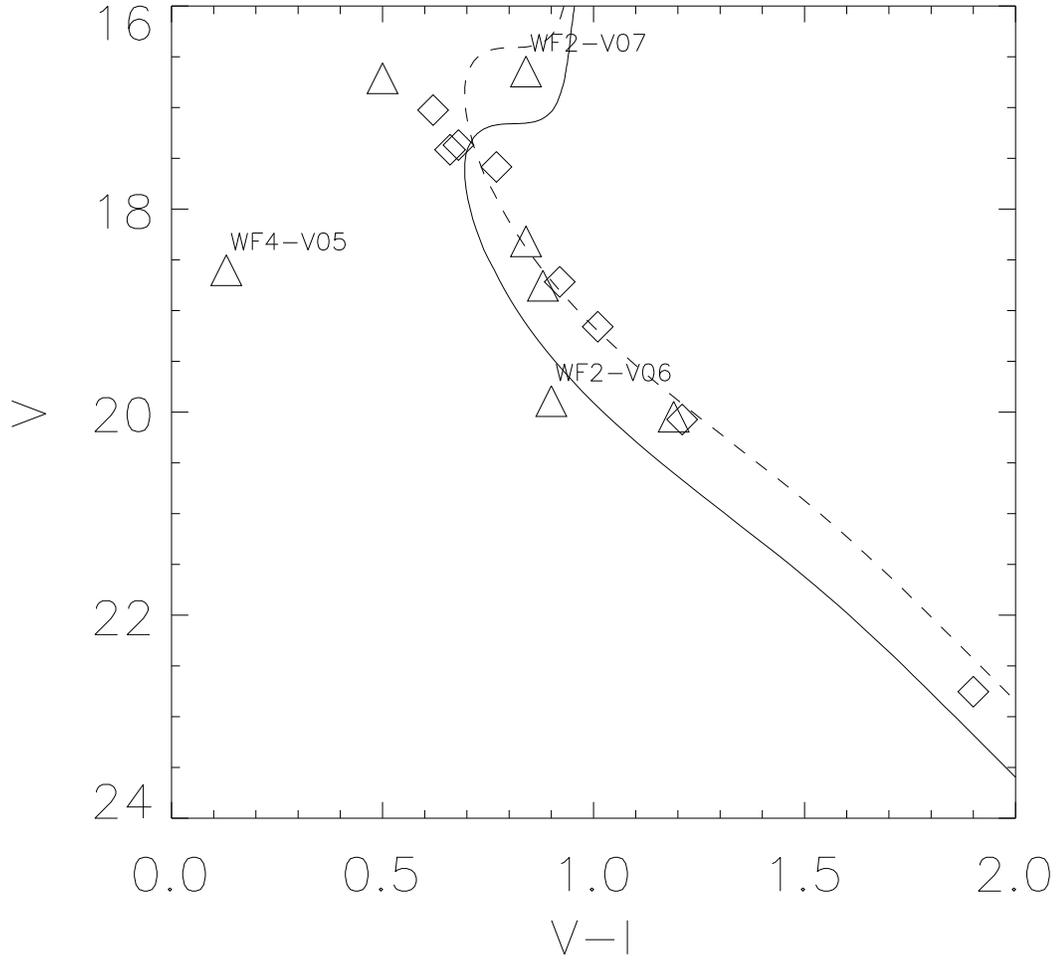}
\caption{\label{CMWUMa} 
Color-magnitude diagram of the W UMa variables. The solid line is the
fiducial main sequence ridge line of all the photometered stars.  The
dashed line is this line shifted upwards by 0.75 mag, i.e. the
equal-mass binary main sequence. Symbols are as in Fig.~\ref{DMWUMa}}
\end{figure}

\begin{figure}
\epsscale{1.0}
\plotone{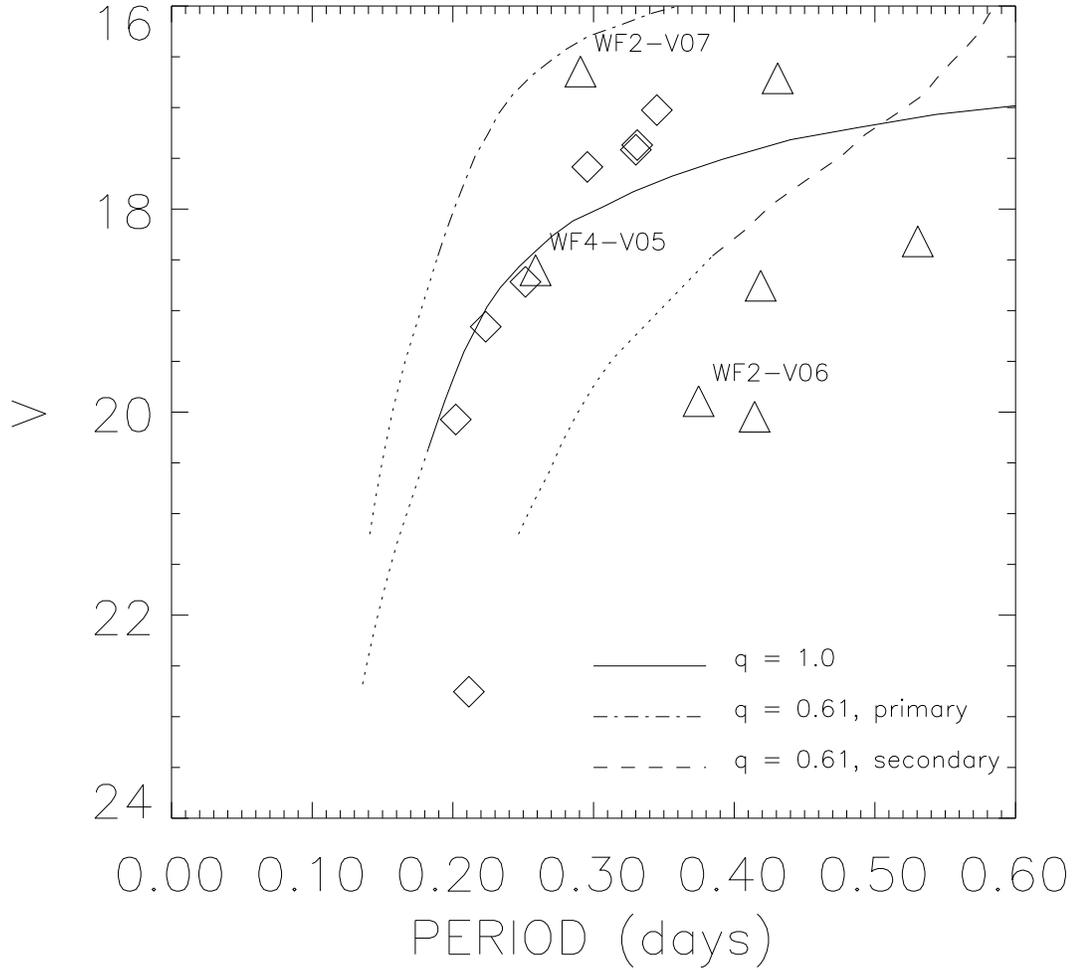}
\caption{\label{PLWUMa} 
The period-luminosity diagram for the W UMa variables. Shown are lines
representing equal-mass Roche-lobe filling and primary and secondary
Roche-lobe filling for a binary with mass ratio $q = 0.61$. Dotted parts of
the curves were calculated using isochrones we have extrapolated to lower 
mass. Symbols are as in Fig.~\ref{DMWUMa}}
\end{figure}

As shown by \citet{Rucinski94,Rucinski95}, contact binaries are observed 
to obey a
period-luminosity-color-metallicity relation,
\begin{equation}
\label{DMeqn}
M_{V}= -4.43\log P+3.63(V-I)_{0}-0.31-0.12[Fe/H],
\end{equation}
with P in days and $M_{V}$ magnitude at maximum.
In Fig.~\ref{DMWUMa} we show the apparent distance modulus of the W
UMa stars as a function of their periods, where we have used
Eqn.~\ref{DMeqn} and followed \citet{Kaluzny98a} in assuming $[Fe/H] =
-0.76$ and $E(V-I) = 0.05$ \citep{Harris96}. For reference, the
apparent distance modulus of 47 Tuc listed in \citet{Harris96} is
$(m-M)_{V} = 13.37$ while \citet{Zoccali01} found $(m-M)_{V} = 13.27$.
In the diagram, we have grouped the stars into those that are
consistent with the Rucinski calibration (Group 1) and those that are
not (Group 2), plotted as different symbols.  The criterion we used
for Group 1 membership was that a star had an apparent distance
modulus within 0.3 magnitudes of 13.1 in Fig.~\ref{DMWUMa}.  We retain
these groupings and plot the color-magnitude diagram
(Fig~\ref{CMWUMa}) and period-luminosity diagram (Fig~\ref{PLWUMa})
for the same stars. In Fig.~\ref{PLWUMa} we also show fiducial lines
representing equal-mass Roche-lobe filling and primary and secondary
Roche-lobe filling for a binary with mass ratio $q = 0.61$.  These
lines were calculated using the condition that the radius of each
component star is equal to the volume-averaged radius of its Roche
lobe \citep{Paczynski71,Eggleton83},
\begin{equation}
R_{L} = \frac{0.49 q^{2/3} a}{0.6 q^{2/3} + \ln(1 + q^{1/3})},
\end{equation}
where $q$ is the mass ratio and $a$ is the separation. The orbital
period follows from Kepler's third law, taking the main-sequence
masses and radii from the Z=0.004 Padova isochrone
\citep{Bertelli94}. (We use here the Padova isochrone because it extends 
further down the main sequence than the Vandenberg isochrone used in 
\S \ref{EBsect}.)
The 11 Gy isochrone was used for the secondary component but for the
primary-star we used the 1 Gy isochrone.  This is appropriate because,
assuming the binaries have not been in contact for a time comparable
to the cluster age, mass transferred to the primary will not have
caused it to evolve off the main sequence (i.e. it will appear in a
retarded evolutionary state relative to single stars in the cluster
with the same mass).

The Group 1 stars all lie on or very close to the binary main sequence
in Fig~\ref{CMWUMa} or are blue stragglers. In Fig.~\ref{PLWUMa} these
stars are found on or very close to the equal-mass main-sequence
critical contact line or (for the brightest few stars) are consistent
with being secondary-Roche-lobe filled binaries with mass ratios $\sim
0.8$.  These stars are normal contact binaries.

The stars in Group 2 do not obey the Rucinski calibration and require
other explanations. The star that deviates the most from the
calibrated absolute magnitude (WF4-V05) lies on the equal-mass
critical contact line in Fig.~\ref{PLWUMa} but is found far to the
blue of the main sequence in Fig~\ref{CMWUMa}. The lightcurve
(Fig.~\ref{lcWUMa1}) appears to have unequal eclipse maxima indicating
a semi-detached system. The very blue colors may indicate mass
transfer from the secondary causing a hot spot on the surface of the
primary or a surrounding accretion disk. The star may be a cataclysmic
variable.

The star in Fig.~\ref{DMWUMa} with the next-highest positive 
deviation (WF2-V06) also has
rather blue colors but this time is found on the long period side of
the Roche-lobe-filling lines in Fig.~\ref{PLWUMa}. This star may be a
semi-detached low-mass ratio binary undergoing mass transfer.

The star which falls below the Rucinski calibration in
Fig.~\ref{DMWUMa} (WF2-V07) is a subgiant. In Fig.~\ref{PLWUMa}, it
lies close to the line of primary Roche-lobe filling for $q = 0.61$,
so should be a contact binary if $q \ga 0.6$. Since it deviates
strongly from the Rucinski calibration, it may have a low mass
companion and be a semi-detached system.

The four remaining Group 2 stars have longer periods than the
secondary-Roche-lobe-filling condition for $q = 0.61$ (Fig.~\ref{PLWUMa})
and lie below or just on the equal-mass binary main sequence
(Fig.~\ref{CMWUMa}). These are most probably semi-detached
systems.

We can use our Group 1 stars to re-derive a distance modulus to 47 Tuc
based on the Rucinski calibration. The mean distance modulus of our 8
Group 1 stars is $(m-M)_{V} = 13.10 \pm 0.15 \pm 0.23$ where the
quoted uncertainties are random followed by systematic. (We have
assumed systematic uncertainties of 0.05 mag in $V$ and in $V-I$.  A
reduction in these with more definitive absolute photometry,
particularly for $V-I$, could significantly lower the systematic
error on the distance modulus.)  This can be compared with the recent
\citet{Zoccali01} value $(m-M)_{V} = 13.27 \pm 0.14$ based on the
position of the white dwarf cooling sequence. We note that
\citet{Zoccali01} used $E(B-V) = 0.055$ which, with $A_{V} = 3.2E(B-V)$
and $A_{I}/A_{V} = 0.6$ \citep{Schlegel98},
corresponds to $E(V-I) = 0.072$. If we use this value,
rather than $E(V-I) = 0.05$ \citep{Harris96}, then we derive
$(m-M)_{V} = 13.17 \pm 0.15$, in agreement with
\citet{Zoccali01}. The W UMa-derived distance modulus favors a distance 
at least as low as the white dwarf-determined value. 

The timescale for initially-detached binaries to evolve to contact is
rather poorly known. The model of \citet{Bradstreet94} predicts that
binaries with primordial component masses $\sim$ 0.7--0.9 $M_{\odot}$
and mass ratios greater than $\sim$ 0.7 will have evolved to contact
in a cluster-age time if their initial orbital periods were $\sim$
2--4 days. (Values of $q$ near unity and lower masses give the longest
initial orbital periods.)  Binaries with periods longer than this will
not have undergone a significant change in their orbital period due to
magnetic braking over the cluster lifetime.  We assume the lowest
primordial orbital period is 2.5 days and the highest that could have
evolved in the age of 47 Tuc to contact is 4 days.

After contact is established, estimates of the timescale for a binary
to merge and become a single star are $\sim 10^{9 \pm 1}$ years
\citep{Vilhu82,Rahunen81}. Out of the 46422 stars we have observed, we
have estimated that 15 are the contact or near-contact products of
angular-momentum-loss evolution. The {\em observed} W UMa frequency 
of our sample is therefore 0.032 \%. Because of non-uniform area
coverage, our survey is skewed towards preferential detection
of variables within 10 arcseconds of the cluster center (see 
\S~\ref{sectObservations}). We can correct for this bias by weighting
each star by the inverse of the radial area completeness function 
(Fig~\ref{RadialArea}), which we use out to a radial distance of 90 
arcseconds. Applying this correction, we have
22.6 W UMa stars detected out of a total of 67004.6 stars. 
Thus the true {\em observed}
W UMa frequency within 90 arcseconds of the cluster center is
0.033 \%,  
one per 3010 stars. This is an order of magnitude
lower than the rate of one W UMa per 250-300 main-sequence stars in
the disk of the Galaxy \citep{Rucinski97}. It is also about 1.5 times
higher than (but probably in statistical agreement with) the observed
main-sequence W UMa frequency in the previous HST survey of the core
of 47 Tuc by
\citet{Edmonds96}. Our observed frequency is 
three times higher than the frequency observed by \citet{Kaluzny98a}
over a large area surrounding but outside of the core. This result is
not unexpected since there is an appreciable mass segregation in 47
Tucanae (see \S \ref{sectBYDra}).
 
Doubling the observed number of W UMa's to allow for unfavorable
orientations for eclipses and adopting $10^{9}$ years for the
contact-binary lifetime requires that 1 new contact binary be formed
every $2.2{\times}10^7$ years from our sample. Over the 11 Gy lifetime
of the cluster, $\sim$ 490 such contact systems would have formed.  If
all the binary systems with primordial periods between 2.5 and 4 days
have evolved to contact or near-contact then our numbers imply a
primordial binary frequency of 0.73\% for those with initial orbital
periods between 2.5 and 4 days.  If we make the further assumption (as
in \S \ref{EBsect}) of a flat distribution of primordial binaries with
$\log P$, this implies a total primordial binary fraction of 14 $\pm$
4 \% (where the quoted error is the lower limit from Poisson
statistics).  This result is in remarkable (but likely fortuitous)
agreement with the value derived from longer period eclipsing binaries
in \S \ref{EBsect}. We stress that the binary frequency derived here
is inversely proportional to the adopted contact binary lifetime. If
the lifetime is $10^{10}$ years rather than $10^{9}$ years then the
derived binary frequency becomes 1.4 $\pm$ 0.4 \%. 

In order to compare 
our value with that for field stars in the solar neighbourhood, we adopt
both a field-star binary frequency (65 \%) and the 'exponential' period
distribution from \citet{Duquennoy91}. Considering only periods between
2.5 days and 50 years, this binary frequency among solar-like stars
is 26 \%.
Thus, given the
assumptions that entered our calculations, the binary frequency in the
core of 47 Tuc (within 90 arcseconds of the center) is about half
the frequency among field solar-type stars.

\section{\label{Shortsect} Other short-period variables}

In addition to those binaries discussed above, we have found a further
15 photometric variables with doubled periods (i.e. twice the observed
periods) of between
0.1 and 1.5 days. Classification of these stars as contact or
semi-detached binaries or BY Dra stars is not possible on the basis of
lightcurve morphology (in the majority of cases) because the amplitude
of variability is too low. Some of them (particularly with periods
less than 0.6 d) may be either contact or semi-detached binary systems
where the orbital inclination is low relative to the plane of the
sky. In these stars the orbital period will be twice the period we
detect.

Others have clearly non-sinusoidal lightcurves. These stars are
unlikely to be pulsators since they are not found in any particular
variable-star region of the color-magnitude diagram.  From
\citet{Hesser87}, we estimate that at most 2 stars in our sample could
be SMC horizontal-branch stars (most of which would be non-variable)
and these would appear to the blue of the 47 Tuc main sequence at $V =
19.4$.  Our provisional interpretation is that they are most likely BY
Draconis variables.  These are defined as low amplitude variables with
periods of a few days where the source of variability is rotational
modulation by starspots \citep{Bopp73}. 
In these cases the rotation period is the same as the photometric period.
Amplitudes of photometric
variability are typically a few hundredths to a few tenths of a
magnitude in $V$
\citep{Cutispoto93} although there is probably a strong selection 
effect against discovery of lower amplitude variables.  Spectral
observations of nearby BY Dra stars imply that many but not all of
them are binaries \citep{Bopp77} but that rapid rotation is the
condition necessary for the BY Dra phenomenon \citep{Vogt83}.  Since
the spin-down time for primordial rotating single stars is small
compared to the age of 47 Tuc
\citep{Skumanich72,Soderblom83,Siess97}, the stars in our sample 
are also likely to be tidally coupled binaries.

\begin{figure}
\epsscale{1.0}
\plotone{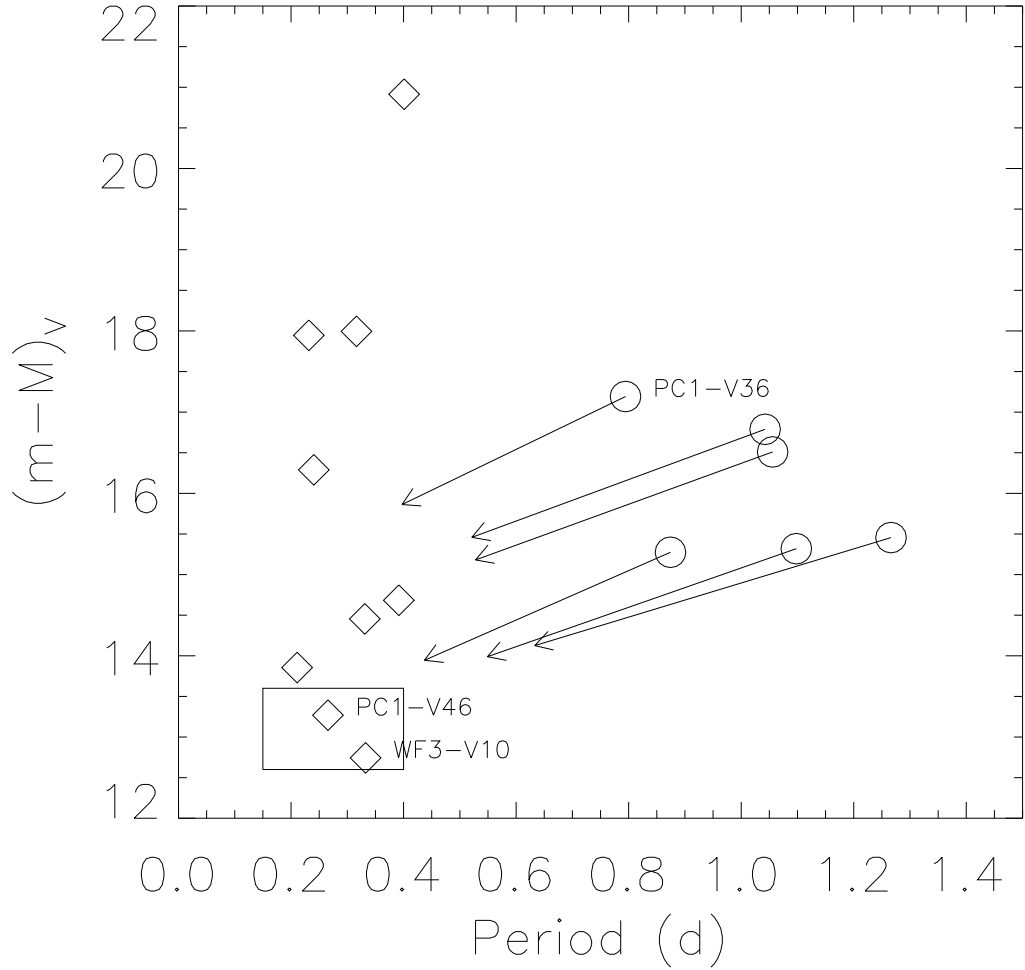}
\caption{\label{DMshort} 
Distance modulus of the short-period non-eclipsing variables according
to the calibration of \citet{Rucinski95}. Stars are plotted assuming their
rotation periods are twice the observed periods. Probable contact or 
semi-detached systems are plotted as diamonds. Circles represent a 
longer-period group of stars which are probably not in or close to contact. 
Arrows indicate the positions
of the longer period group if their observed periods are the periods
of rotation, in which case near-contact conditions hold.}
\end{figure}

\begin{figure}
\epsscale{1.0}
\plotone{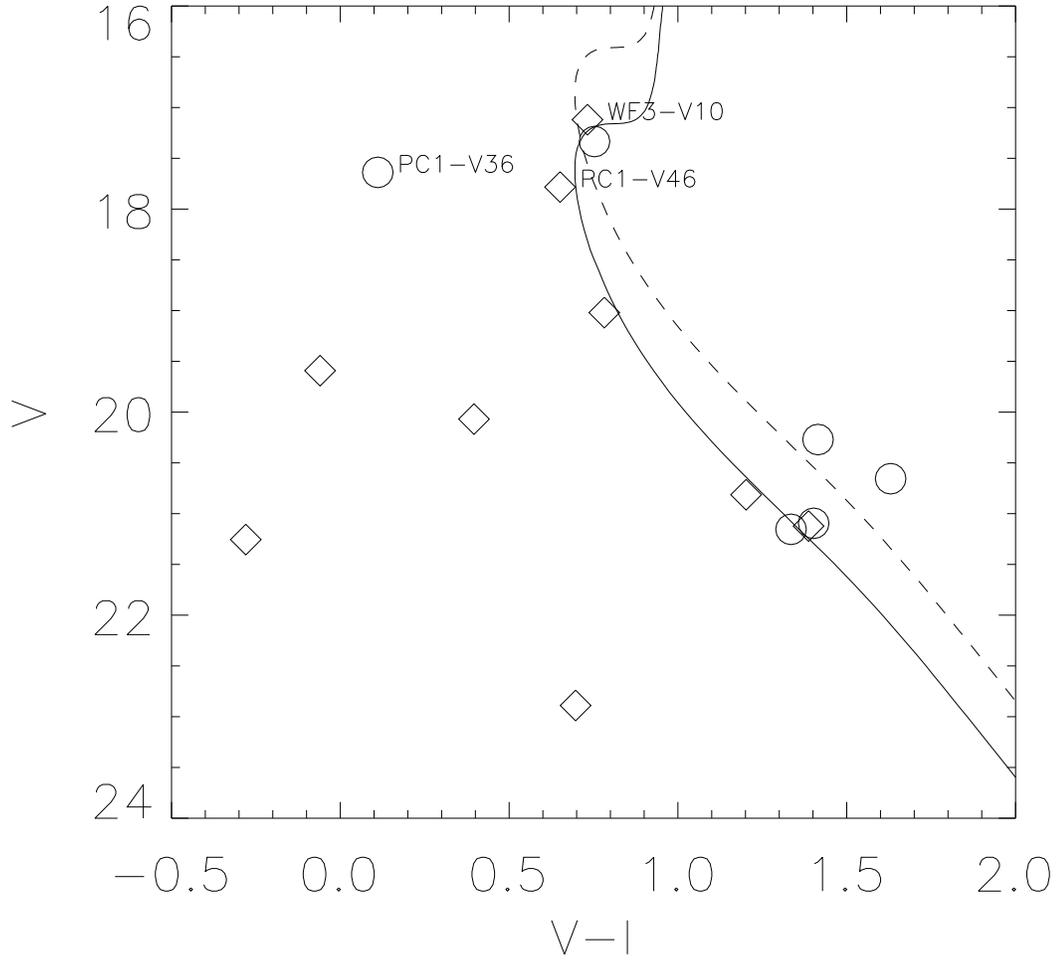}
\caption{\label{CMshort} 
Color-magnitude diagram of the short-period non-eclipsing
variables. The solid line is the fiducial ridge line for 47 Tuc.  
The dashed line is this line shifted
upwards by 0.75 mag, i.e. the equal-mass binary sequence.
Symbols are as in Fig.~\ref{DMshort}.}
\end{figure}

\begin{figure}
\epsscale{1.0}
\plotone{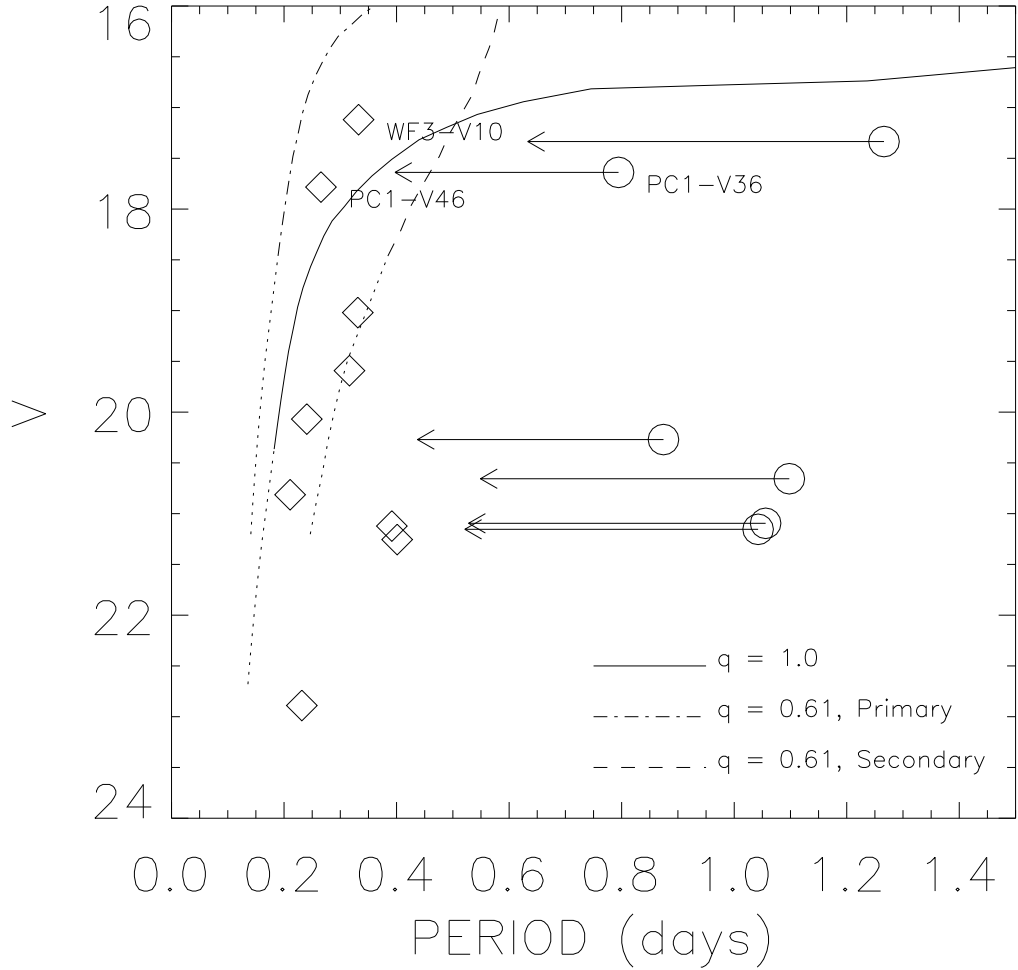}
\caption{\label{PLshort}
The period-luminosity diagram for the short-period non-eclipsing
variables. Shown are lines representing equal-mass Roche-lobe filling
and primary and secondary Roche-lobe filling for a binary with mass
ratio $q = 0.61$. Symbols are as in Fig.~\ref{DMshort}.
Arrows show possible locations for cases
in which a factor of two rotation-period abiguity remains.}
\end{figure}

As mentioned earlier, 
classification of most of these low-amplitude variables from their lightcurve 
morphology is not possible. To ask which of them are 
consistent with being contact
binaries, we again turn to the absolute magnitude calibration of
\citet{Rucinski95} (Fig.~\ref{DMshort}) and the period--luminosity
(Fig.~\ref{PLshort}) and color--magnitude (Fig.~\ref{CMshort})
diagrams.

In Fig.~\ref{DMshort} we indicate the box where the contact W Uma
binaries were found in \S \ref{WUMasect} and divide the stars
into two groups based on their (double) periods.  There is an apparent period
gap between 0.4 and 0.8 days and we hypothesize that the double period
is incorrect for the longer period group. From Fig.~\ref{PLshort},
even the single periods of these stars are generally too long for
secondary Roche lobe filling and their absolute magnitudes are
inconsistent with the Rucinski calibration for contact binaries. In
Fig.~\ref{CMshort}, this group, with one exception, lie on or near the
single or binary main sequences. These are most likely BY Draconis
stars. We note overlap between these stars 
and the longer period variables considered in \S \ref{sectBYDra}.

The properties of the one member of this group with very blue colors
(PC1-V36) is rather puzzling. In the period-luminosity diagram
Fig.~\ref{PLshort} it lies in the region where secondary Roche lobe
filling should have occured if the period we observe (0.4 d) is the
rotation period, not half the rotation period. If the Roche lobe has
been filled however, we should observe ellipsoidal variability at half
the rotation period. On the other hand, if the rotation period is 0.8 d
then we would not expect Roche lobe filling and the consequential
ellipsoidal variability. However, if the secondary star in this system
is of very low mass, then Roche lobe filling may still have occurred, even
at a period of 0.8 d. The blue $V-I$ colour is indicative of ongoing
mass transfer.  We suggest this as a likely scenario for this star,
and hence that the true rotation period is 0.8d. If this is the case
then the star belongs with the short-period group discussed next and is
not a BY Dra star.

The short-period group of stars are found in Fig.~\ref{DMshort} in the
same region or extending above the W UMa binaries in
Fig.~\ref{DMWUMa}. Stars PC1-V46 and WF3-V10 are consistent with the
Rucinski calibration and have periods and luminosities in
Fig.~\ref{PLshort} consistent with having filled their primary and
secondary Roche lobes for $q \sim 0.8$. These are likely to be contact
binaries. (We elect to keep them in a separate group from the W UMa
stars since our classification is not on the basis of lightcurve
morphology.)  The remaining stars have blue colors and have periods
consistent with being close to secondary Roche lobe filling. These are
probably the non-eclipsing counterparts of the mass-transferring
semi-detached binaries found in \S \ref{WUMasect}.  The observed parameters
of
these non-eclipsing contact or semi-detached binaries are listed in
Table~\ref{tableShort}.

\section{\label{sectBYDra} Longer-period variables: the BY Dra stars}

\begin{figure}
\epsscale{0.7}
\plotone{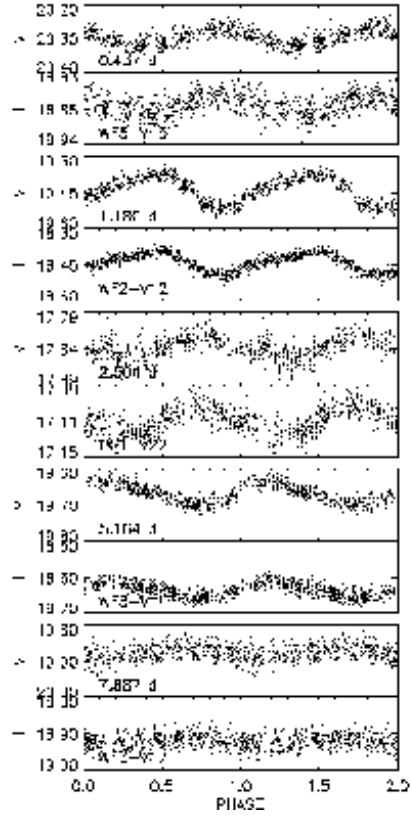}
\caption{\label{lcBYDra} 
Lightcurves of sample BY Dra stars.}
\end{figure}

Additional to the stars classified as BY Dra in \S \ref{Shortsect} and
extending to periods as long as 10 days, we have found a further 69
variables which we also suspect to be of this type.  Sample
lightcurves for a representative range of periods are shown in
Fig.~\ref{lcBYDra}. Several of the variables are of very low amplitude
and are only recognized as such through their power spectra.

The color--magnitude diagram for these stars (and also those thought
to be BY Dra type in \S \ref{Shortsect}) is plotted in
Fig.~\ref{CMBYDra}. The majority of stars are within the band defined
by single and binary main sequences (Table~\ref{tableBYDra}).  Notable
are a group of six red stars in a box slightly below subgiant
magnitudes at $V \: \sim \: 17.5$ (Table~\ref{tableRS}, which we
discuss further in \S \ref{RSsect}), two blue stragglers and
three fainter stars with very blue $V-I$ color
(Table~\ref{tableOther}).

A powerful test of the binary hypothesis for this group of stars is to
compare their radial distribution with that of the main-sequence stars
in the cluster, the blue straggler stars (BSS) and that of the
contact, semi-detached and eclipsing binaries. The cumulative
distributions for these samples are shown in Fig.~\ref{CFD} where we
have excluded the two blue stragglers, the three very blue variables
and the six red stars from the BY Dra sample.  (The faint blue
variables may be candidates for cataclysmic variables and will be
considered in a separate study by Edmonds {\em et al.} 2001, in
preparation.) 
These distributions have been corrected for radial area incompleteness
using Fig~\ref{RadialArea}.
A shoulder in the distributions near a radial distance
of 20 arcsec (which was stronger before the radial area coverage
correction) is still noticeable. This is  due
to a higher level of crowding (relative to pixel size) in the
WF CCD's than in the PC at the same radial distance from the 
center of the cluster.
As noted in \S~\ref{sectObservations}, our master star
list does not contain any stars where more than 90 \% of the light
within an aperture of radius 5 (PC) or 4 (WF) pixels centered
on the source comes from neighbouring stars. Such a bias will preferentially
act against fainter stars being included. We have made no attempt
to correct for this effect here.

To emphasize the degree of mass segregation in the
cluster, we have split the main-sequence stars into three groups; $17.0
\leq V < 18.5$, $18.5 \leq V < 20$, $20 \leq V < 21.5$.  The
main-sequence stars show an increasing degree of central concentration
with brightness, perfectly illustrating the mass-segregation effect.
(We have not included corrections for incompleteness as a function of
radius and magnitude, but expect such to be small even for the last
main-sequence bin which extends only to $V$ of 21.5; these details
will be explored in Guhathakurta {\em et al.} 2001, in preparation.)
The cumulative distributions for the suspected BY Dra stars and the
other binaries are very similar and show a higher degree of central
concentration than the most massive main sequence stars.  The blue
stragglers are the most centrally concentrated of all and are likely
to be close binaries or their merged remnants.  Since the majority of
blue stragglers are found more than 0.75 magnitudes above the
main-sequence turnoff, at least one of the component masses in these
systems is likely greater than the current main-sequence turn-off
mass.

To formalize the differences between the distributions, we have used
the Kolmogorov-Smirnov test.  Applied to the BY Dra and binary
samples, the KS statistic (the maximum difference between
distributions) is 0.13 with a 85.5\% probability that they are drawn
from the same population.  The difference between the BY Dra and
brightest main-sequence group is slightly greater (KS statistic 0.15) with only
a 12.5\% probability that they are drawn from the same population.  This
is convincing evidence that our suspected BY Dra stars are indeed
binaries.

From the distributions Fig.~\ref{CFD}, we can investigate 
(in a broad sense) how
the binary fraction of the cluster changes as a function of radius.
Combining the BY Dra and ``Binary'' distributions, we have compared the
fraction of detections in three radius bins,
$0 \leq r < 1$, $1 \leq r < 2$ and $2 \leq r < 3$,
where $r$ is the radial distance from the cluster center in units
of 24 arcseconds, the core radius \citep{Howell00}.
Scaled to match an overall binary frequency of 13.5 \% 
within 90 arcseconds (an average from \S \ref{EBsect},
\ref{WUMasect}), the binary frequency in the three radial bins is
20.3 \%, 18.3 \% and 7.7 \%, indicating a relatively constant fraction
within two core radii, decreasing rapidly outside of that.

Since the BY Dra stars are presumably the non-eclipsing counterparts
of the detached eclipsing binaries discussed in \S \ref{EBsect},
we expect that their amplitudes of variability should be consistent
with those of the eclipsing binaries outside of eclipses (that is, the
O'Connell effect).  In Fig.~\ref{PABYDra} we plot the semi-amplitudes
of variation for these two groups as a function of period, again
excluding the blue stragglers and fainter blue variables.  As a
separate grouping, we include the six red variables from the box 
in Fig.~\ref{CMBYDra}.  There is
a tendency for variability amplitude to decrease with increasing
period.  Overall, the distribution of amplitudes is consistent between
the BY Dra stars and the eclipsing binaries. Two of the six red variables
have amplitudes much greater than the eclipsing binaries.

The large population of BY Dra variables was unexpected and deserves
further comment both as a general class and for what they imply for 
binary fraction inferences, and also for an interesting sub-grouping
of these within the CMD.
In order to have enhanced activity we rely on the assumption that 
the stellar rotation rate is rapid due to synchronous rotation with the
binary orbit.
The radial concentration of this class of objects (see Fig.~\ref{CFD})
supports the assumption that these are binaries.
For circular orbits the rotation and binary periods would be equal.
If the orbit is eccentric, then the rotation period may be much shorter
than the binary period, as driven by the much higher orbital velocity
at periastron.
The binary S1242 in M67 has $e$ = 0.66 and an orbital period of 31.8
days \citep{VandenBerg99} based on radial velocities.  \citet{Gilliland91}
reported a photometric period of 4.88 days at 0.0025 mag amplitude.
\citet{VandenBerg99} note that the photometric period corresponds to
corotation with the orbit at periastron.  With the synchronization
timescale scaling as $(a/R_*)^6$ \citep{Keppens97} it is easy to see
how such a system can arise.

The timescale for rotational synchronization should be short enough to
enforce rapid rotation for all binaries with $P \: \lesssim$ 10 days
(or equivalent periastron values), even for systems significantly younger
than the age of 47 Tuc, but the theory is complex with uncertainties
remaining \citep{Keppens97,Keppens00}.
We may reasonably assume that any binaries with periods less than or
comparable to our sample length of 8.3 days will have synchronously 
rotating stars.
\citet{Vogt83} surveyed BY Dra stars and found that whether single or
in binaries, rotation was more rapid than for other dwarfs and
concluded that 5 $km~s^{-1}$ or faster rotation was the underlying
cause of BY Dra activity.  A rotational velocity of 5 $km~s^{-1}$
corresponds to $P_{rot} \: \sim$ 8 days for a 47 Tuc K dwarf with $V
\: \sim$ 19.5.  The general decline of amplitude with period shown in
Fig. 20 is expected, but our recovery rate for BY Dra stars in the
cluster depends on the (unknown) underlying distribution of BY Dra
star amplitudes.

We may use the BY Dra variables to obtain a lower limit to the binary
fraction. We have detected 31 BY Dra stars and 5 eclipsing variables
with periods longer than 4 days. If we assume that these represent all
the binary stars in our sample in this period range and again make the
assumptions of a flat primordial binary distribution with $\log P$
truncated at 2.5 days and 50 years then (after a slight adjustment for 
area-incompleteness) the total binary fraction is
0.76 \%. This is a factor 17 lower than the estimates from the numbers
of eclipsing binaries and W UMa stars.  In Fig.~\ref{PABYDra}, a
significant number of BY Dra stars are detected with millimagnitude
amplitudes in $V$ and there may well be large numbers at levels
undetectable in our data. From Fig~\ref{RecoveryRate}, our theoretical
detection efficiency drops to close to zero for stars fainter than $V
= 18.5$ and $V$--amplitudes of 5~mmag. For 1~mmag amplitudes the loss
of detectability occurs at $V \sim 17$. Thus, the observed BY Dra
frequency is probably low because of incompleteness of the sample,
but it is also possible that low-amplitude surface activity is a
transient phenomenon in these stars.

The prevalence of BY Dra stars in 47 Tuc emphasizes the importance of
including angular momentum loss from stellar winds ({\em e.g..}
\citet{Keppens00}) in any general consideration of binary evolution
once the period has dropped below about 10 days.  Single stars (e.g. 
Sun at 5 Gyr old) are rapidly braked to rotation periods $>$ 20 days at
which point activity levels drop.  In a binary with forced synchronous
rotation, magnetic braking feeds back on the orbit which leads to a
decreasing orbital period and hence greater activity in response to
increasingly rapid rotation.  The large angular momentum reservoir of
the binary can force high rotation rates for extended periods.
\citet{Soderblom90} noted that field BY Dra stars are a dynamically 
young (1 -- 2 Gyr) population, hence raising the lifetime issue in using
BY Dra stars (or eclipsing binaries) to infer an overall binary fraction.
The median period of BY Dra stars surveyed by \citet{Soderblom90} was 3.0
days, not significantly different from the 47 Tuc sample.

\subsection{\label{RSsect}Red stragglers:  a distinct sub-class?}

Several stars in Fig. 18 fall well away from the primary stellar
sequences (main sequence, binary sequence, subgiants, giants, and blue
stragglers).  The three stars fainter and bluer than the main sequence
may be cataclysmic variables and should not be counted as BY Dra, or
related stars.  The six stars falling within $17.25 \lesssim V \lesssim 17.75$
and well to the red of primary main sequence do not have a ready
explanation. We propose to call variables in this region of the 
color-magnitude diagram red straggler stars. 

Inspection of color-magnitude diagrams for fields off the core of 47
Tuc region \citep{Hesser87,Zoccali01} shows that the background SMC
giant branch intersects the 47 Tuc main sequence at $V = 20$ and
passes through the red straggler region. If the ratio of giants to
main sequence stars is similar in the SMC as for 47 Tuc, then from
\citet{Zoccali01} we might expect 1 SMC giant to appear in our red
straggler box. However, if the red stragglers were truly background
SMC stars then we would expect them to lie in a diagonal sequence in
$V,V-I$, not a near-horizontal grouping.  A Kolmogorov-Smirnov test
of the radial distribution from the cluster center of the five stars
plus PC1-V11 (a known cataclysmic variable found in the same location
in the CM diagram - see below) compared to the radial distribution of
area on the sky covered by the WFPC2 CCDs (SMC stars should be
distributed equally in area) gives only a 12\% probability that they are
drawn from the same distribution.

\begin{figure}
\epsscale{1.0}
\plotone{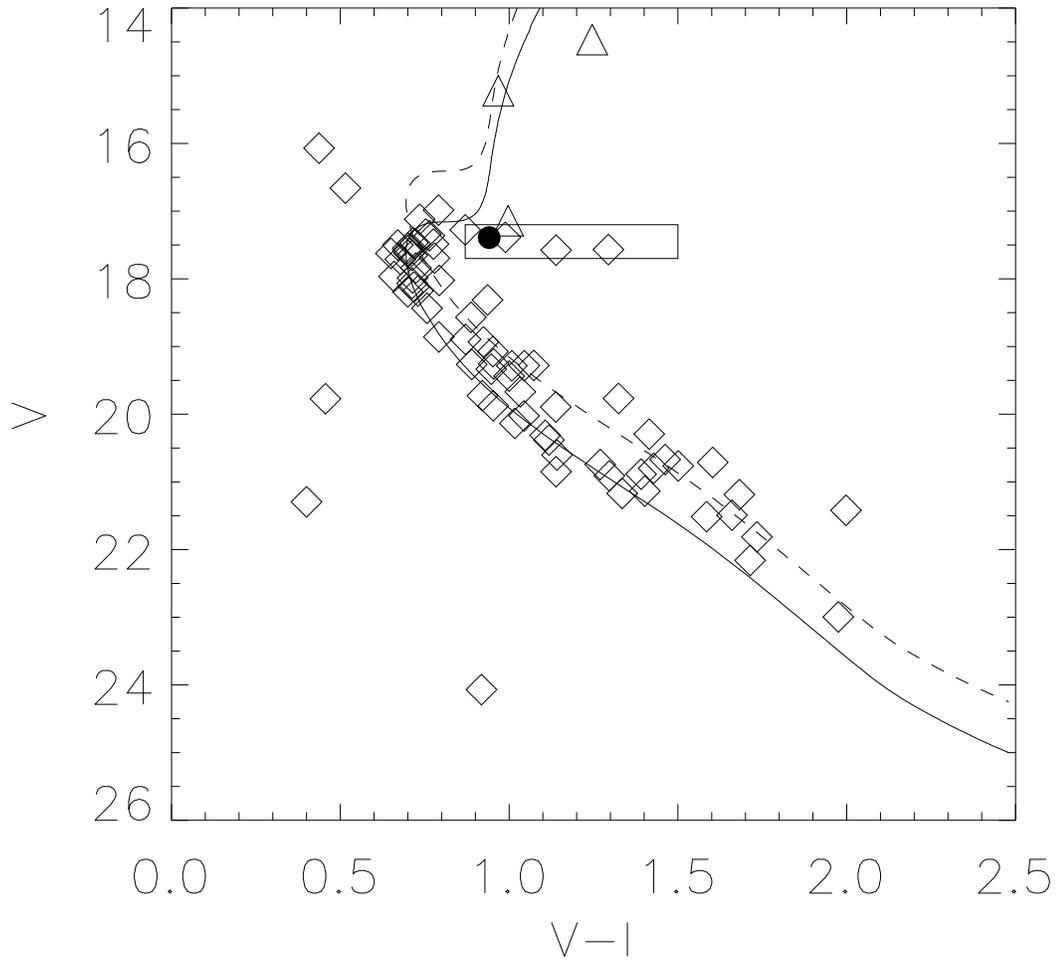}
\caption{\label{CMBYDra} 
Color-magnitude diagram of the suspected BY Dra variables. The solid
line is the fiducial ridge line for 47 Tuc.  The dashed line is this
line shifted upwards by 0.75 mag, i.e. the equal-mass binary main
sequence. Symbols are: diamonds, suspected BY Dra stars; triangles,
variables from saturated-star photometry; filled circle, PC1-V11 (a
known cataclysmic variable).  The box near $V = 17.5$ outlines the
proposed red straggler domain (see text \S~\ref{RSsect}).
}
\end{figure}

\begin{figure}
\epsscale{1.0}
\plotone{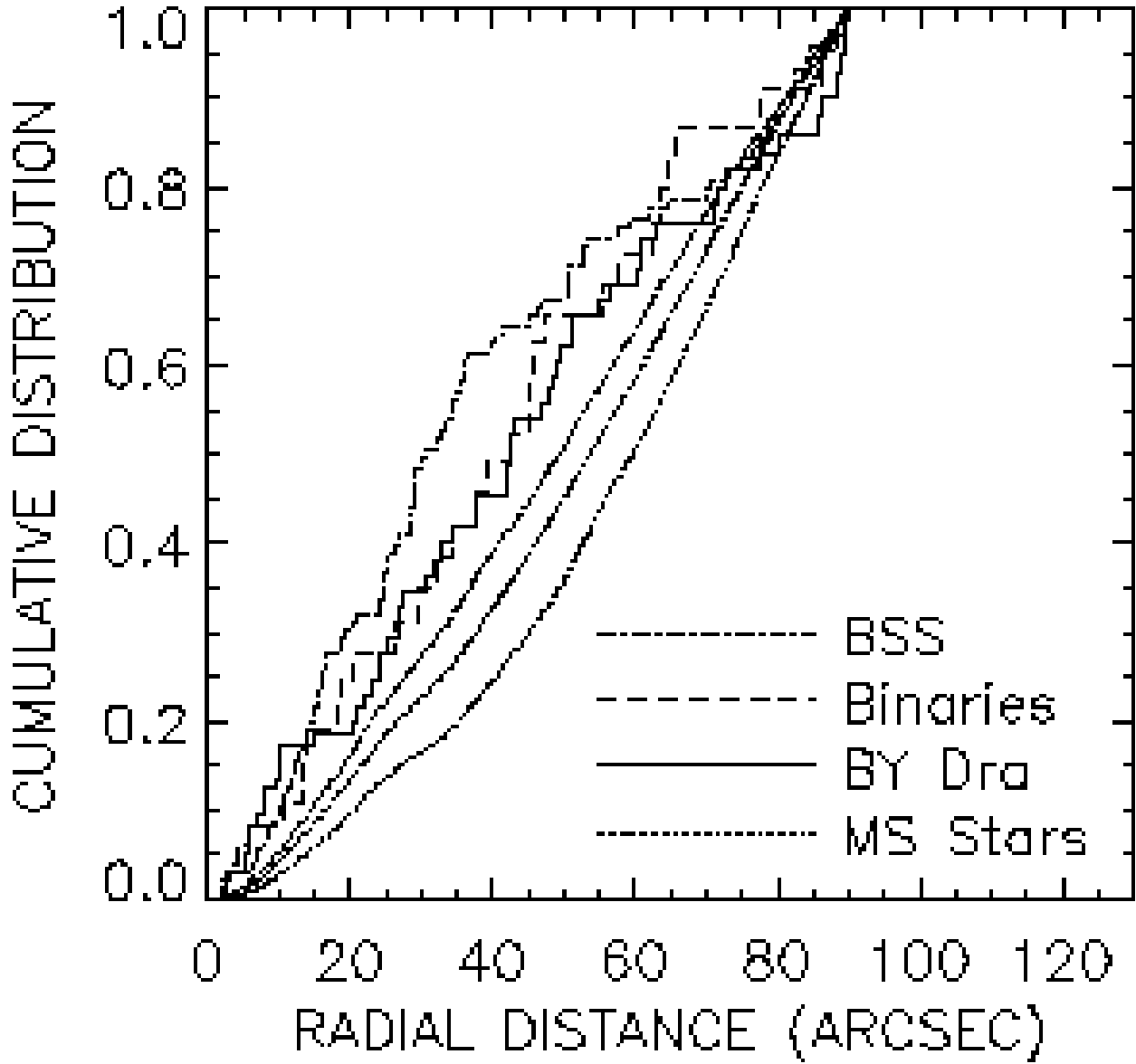}
\caption{\label{CFD} 
Cumulative distribution with cluster radial distance out to 
90 arcseconds of the suspected
BY Dra stars compared with binaries (includes contact, semi-detached
and eclipsing), blue-straggler stars and main sequence stars from 
three successive 1.5 magnitude bins below cluster turnoff.
The highest MS-star distribution is for the brightest magnitude bin
and the lowest is for the faintest bin.
The distributions shown have been corrected for radial incompleteness
in the area coverage of our survey.}
\end{figure}

\begin{figure}
\epsscale{1.0}
\plotone{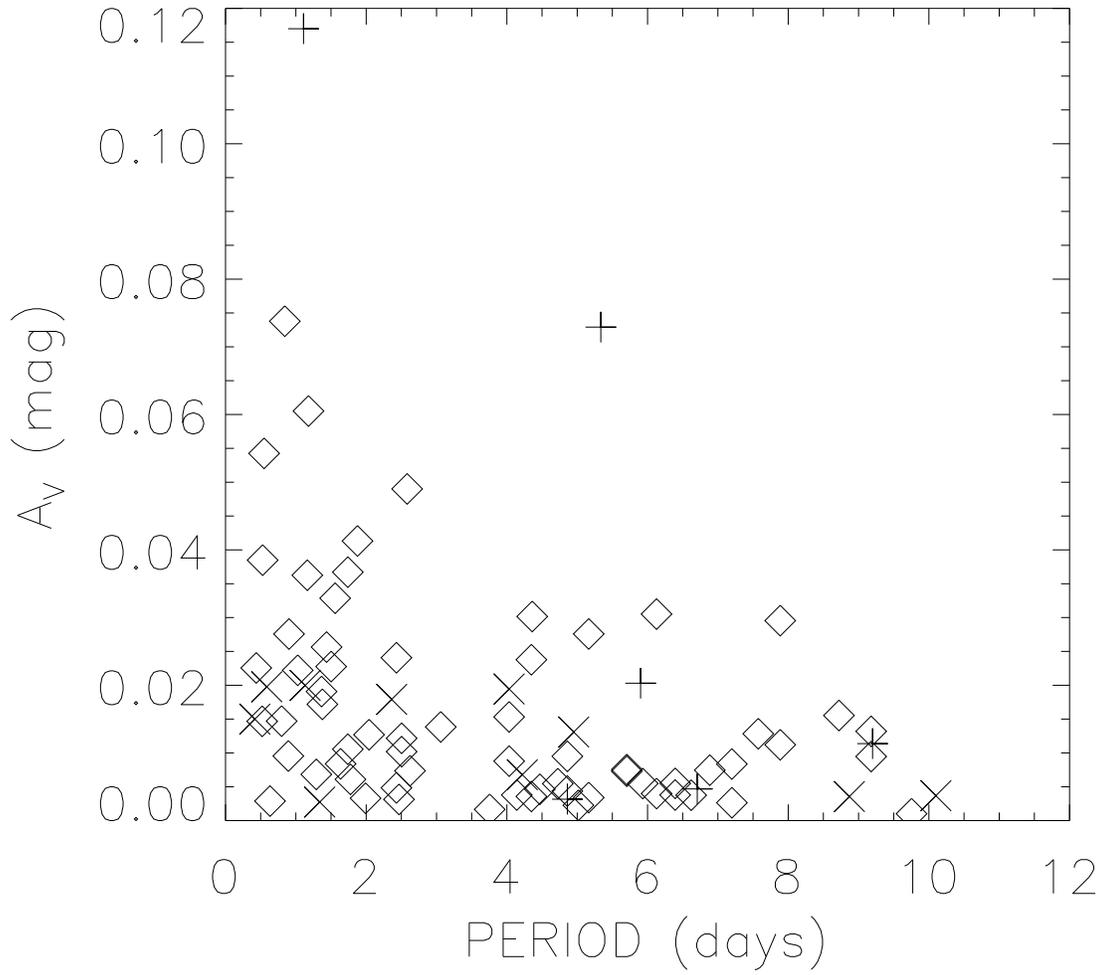}
\caption{\label{PABYDra} 
The distribution of semi-amplitudes of the BY Dra variables (diamonds)
eclipsing binaries outside of eclipse (crosses) and red straggler stars
(plus signs).}
\end{figure}

Noting that (a) these are closest in the CMD to subgiants, and (b)
similarly very red variables do not appear at fainter levels to the
right of the main sequence suggests that their origin is associated
with subgiants or giants.  Two stars (S1063 and S1113) in M67 (a rich
open cluster, but with order 1\% the total population of 47 Tuc) fall
in the red straggler region of the CMD (compensating for relative distance modulii they fall within the small $V$, $V-I$ box in Fig~\ref{CMBYDra})
and have been explicitly
commented on in a number of studies by \citet{Belloni98},
\citet{VandenBerg99}, and \citet{Mathieu00} (where they are referred
to as ``sub-subgiant'' stars).  The two M67 stars in the corresponding
CMD location are: (a) spectroscopic binaries with periods of 2.82 and
18.4 days, (b) X-ray sources, (c) active stars with both Ca II H \& K
emission and photometric variability noted.  All three referenced
studies of these well-observed M67 stars note that no ready
explanation for stars in this region of the CMD exist and that their
origin and evolutionary status remains unknown.
\citet{VandenBerg99} do note that in principle a subgiant or giant can
become underluminous when it transfers mass to a companion, but also
that the high eccentricity of one of the M67 stars (S1063 at 
$e$ = 0.2 and $P$ = 18.4 days) is hard to reconcile with this since
rapid circularization of the orbit is expected during mass transfer.
The range of observational data for the corresponding 47 Tuc 
red stragglers is much
less complete as compared to the M67 cases.

The evidence strongly suggests that these variables and the
corresponding ones in M67 form a distinct class: not only are they
clearly variable, but in a CMD with 40000+ entries, they dominate the
membership in the small CMD box that they occupy. From the photometry
of the non-saturated stars, 5 out of the 17 stars from Fig.~\ref{CMcombined}
which lie in the red straggler box indicated in
Fig~\ref{CMBYDra} are variables. Redward of $V-I = 0.94$, 4 out of
5 are variables.  We posit that a plausible explanation for these
stars is a deflated radius from subgiant or giant origins as the
result of mass transfer initiated by Roche-lobe contact by the evolved
star, for which the secondary has a lower mass.  The subsequent
evolution as the orbit contracts will force a smaller stellar radius.

\citet{Gilliland82} discusses subgiant evolution with mass loss for 
the long-period cataclysmic variable BV Cen - a star in 
an analogous state.
Indeed, the observed $V$ and $V-I$ for BV Cen \citep{Menzies86}
adjusted for distance and differential reddening also place it
within our Fig.~\ref{CMBYDra} red straggler box at a location near
that for PC1-V11 (AKO 9).  The expected lifetime of stars like BV Cen
exceeds 10$^{9}$ years
\citep{Gilliland82}.
We also note that the expanded envelopes of these deflated
subgiants/giants will have longer convective turnover timescales
\citep{Gilliland85} relative to main sequence stars at the same color
or luminosity, and thus smaller Rossby number (rotation period divided
by convective turnover timescale).  This, coupled with enforced
synchronous, rapid (relative to other similar stars) rotation in a
short period binary, provides a ready explanation \citep{Gilliland85}
for heightened activity such as for WF2-V31 which 
is an outlier in the amplitude -- period diagram of Fig.~\ref{PABYDra}.

It is also interesting to note that the 1.108 day period
eclipsing-variable PC1-V11 = AKO 9 \citep{Knigge00}, which has clear
spectroscopic signatures in the UV consistent with a long period CV
({\em i.e.,} the system contains a white dwarf), falls within this
red-straggler domain in a $V$, $V - I$ CMD.  The presence of a white
dwarf primary and accretion disk leads, however, to a much brighter
$U$ magnitude than for the other stars in this $V$, $V - I$ region.
BV Cen, as noted above, falls in this domain in $V$, $V-I$ but, like
AKO 9 , has a strong UV excess.
Perhaps the companions for the other red straggler variables in this
domain are main sequence stars, not white dwarfs, which may
explain their lack of excess UV emission.

At least two of the six red stragglers detected here have also been
detected with Chandra (Edmonds et al., 2001 in preparation), and have
X-ray luminosities similar to those of S1063 and S1113 in M67
\citep{Belloni98} and similar to those of field RS CVns
\citep{Dempsey97}.  One other red straggler in 47 Tuc, lying outside our 
field of
view, has been found using archival HST analysis and this star is also
likely to be a Chandra source (Edmonds et al. 2001).  These
observations strengthen our argument that the red stragglers in 47 Tuc
and M67 form a distinct class with properties explained by mass
transfer, rapid rotation and enhanced activity. We suggest that stars in 
other clusters falling within this red-straggler domain 
(e.g. the two stars within this domain in the high-quality NGC 6752 CMD of
\citet{Rubenstein97}) 
are prime
candidates as variables and X-ray sources.

\section{\label{sectOther}Miscellaneous variables}

\begin{figure}
\epsscale{0.7}
\plotone{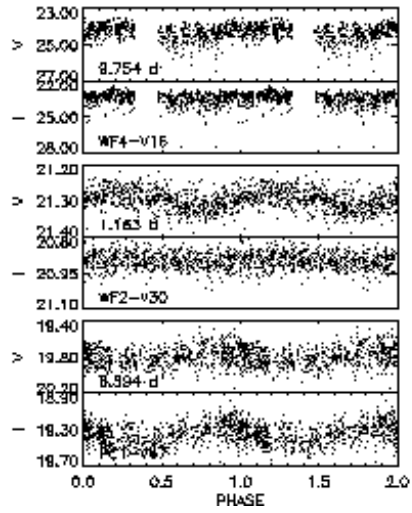}
\caption{\label{lcMisc1} 
Phased lightcurves for three miscellaneous periodic variables.}
\end{figure}

\begin{figure}
\epsscale{0.7}
\plotone{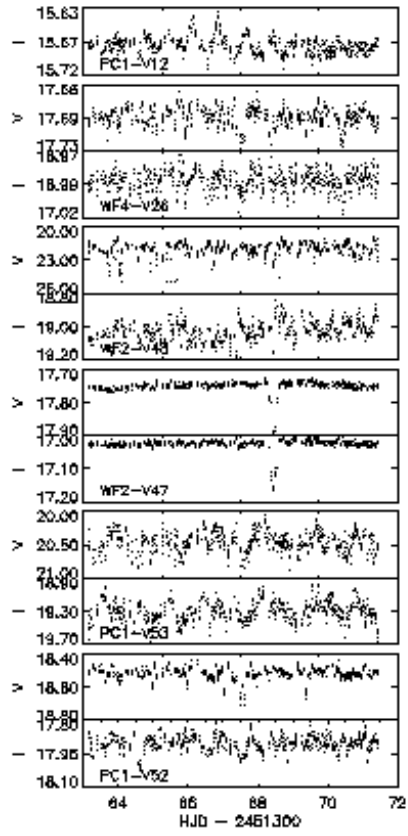}
\caption{\label{lcMisc2} 
Lightcurves for miscellaneous irregular variables.}
\end{figure}

A small number of stars that do not fit into any of the categories
considered were also found to be variable.  These are listed in
Table~\ref{tableOther} and their lightcurves shown in
Figs~\ref{lcMisc1} and \ref{lcMisc2}. The three periodic variables
(PC1-V47, WF2-V30, WF4-V16; Fig~\ref{lcMisc1}) are the blue stars we
excluded from the BY Dra sample in \S \ref{sectBYDra}. These are
candidates for being long period cataclysmic variables.  WF2-V47 is an
eclipsing binary which showed only one eclipse during the time of our
observations. PC1-V12 is a blue straggler whose variability amplitude
seemed to grow and then decline during the time we observed it.
(Unfortunately this star was saturated in $V$ and attempts at $V$-band
photometry were unsuccessful.)  The remaining four stars (PC1-V52,
PC1-V53, WF2-V48, WF4-V26) have irregular lightcurves. PC1-V53 and
WF2-V48 have rather blue $U-V$ colors and WF2-V48 is also unusually
red in $V-I$. We caution that WF2-V48 lies on or very close to
a diffraction spike from a nearby bright star, which may have influenced
the photometry. PC1-V52 lies blueward of the main sequence in
$V-I$. These stars are also candidate cataclysmic variables.

\section{\label{sectGiants}Giant stars}

\begin{figure}
\epsscale{1.0}
\plotone{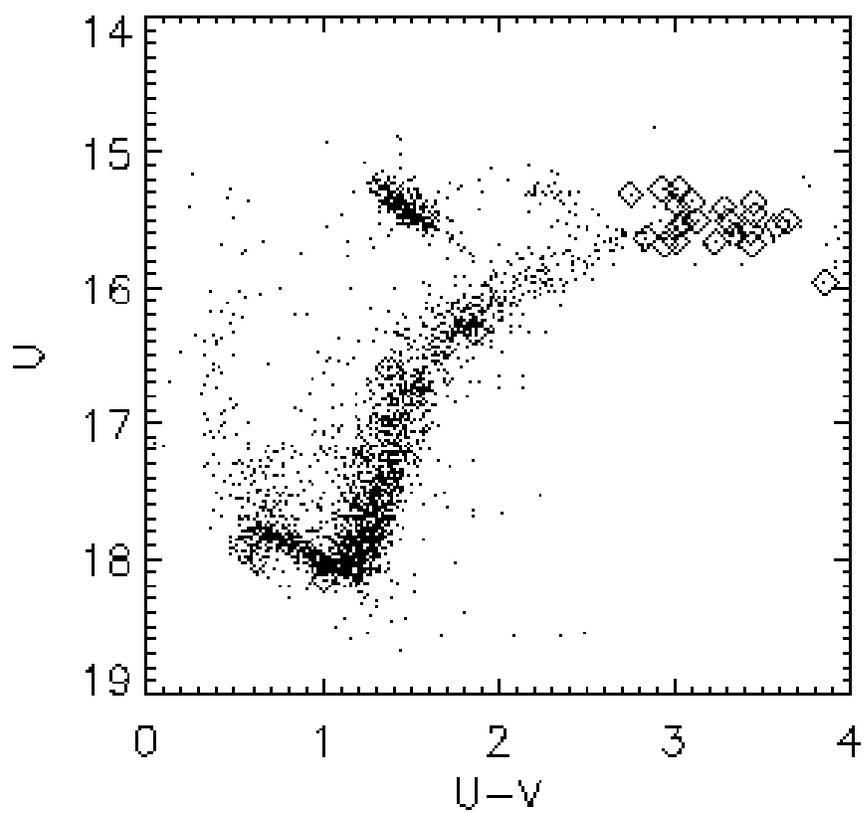}
\caption{\label{CMDSat} 
Color--magnitude diagram from the saturated--star photometry for all
four CCDs with the variable stars indicated.}
\end{figure}

\begin{figure}
\epsscale{0.7}
\plotone{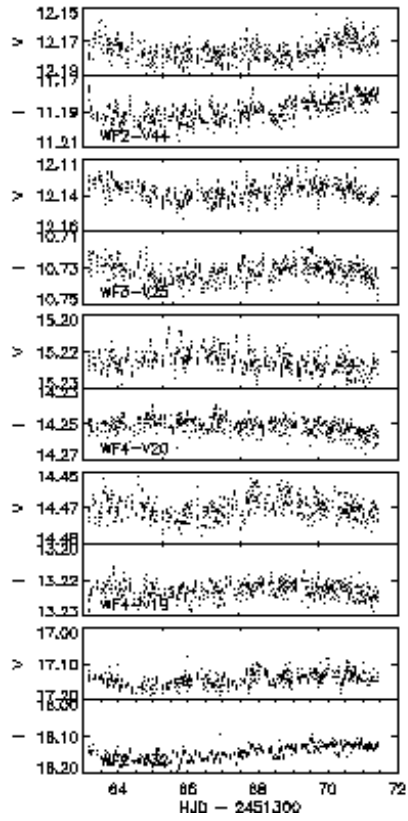}
\caption{\label{lcGiants} 
Sample lightcurves of variable red giants (WF2-V32, WF4-V19, WF4-V20) and
RGB-AGB stars (WF3-V25, WF2-V44) from saturated-star
photometry. }
\end{figure}

Using the methods of \citet{Gilliland94}, time series photometry for
$\sim$3000 saturated stars have been extracted. The quality of this
photometry is not as good (typical errors of 1 -- 3\%) as for the
unsaturated stars and the errors are not likely to be normally
distributed. Nevertheless, we have searched these time series for
stellar variability using the same methods and criteria as for the
unsaturated stars and have found 27 giant stars which appear to be
periodic variables (Tables~\ref{tableRGB} and \ref{tableAGB}).  (The
noise level and its time-correlated non-Gaussian characteristics meant
that the saturated photometry was insensitive to detecting the $\lesssim$
1\% amplitude variables with several-day periods discussed in
\citet{Edmonds96a}.)  The distribution of the identified variables in
the color--magnitude diagram is shown in Fig.~\ref{CMDSat}.  Since the
most strongly saturated stars have extremely poor $I$--band
photometry, we show this diagram in $U$, $U-V$ rather than $V$, $V-I$.
The three least-saturated of these stars are also plotted in 
Fig.~\ref{CMBYDra}.
Two stars, WF4-V19 and WF4-V20 (Table~\ref{tableRGB}, which we label RGB), are giant-branch stars with
$V$--magnitudes fainter than the horizontal branch. They may
be RS CVn systems. One star,
WF2-V32 at $V$ = 17.14, $V - I$ = 0.96, is located in the red straggler
domain discussed in the previous section.   The remaining 24 stars (Table~\ref{tableAGB}, labeled
RGB-AGB) are all $\sim$2 magnitudes brighter in $V$ than the
horizontal branch.  Most of them are probably pulsating semi-regular
variables \citep{Frogel98} and the periods we have found from the
Fourier analysis are likely underestimates.  Lightcurves are shown
in Fig.~\ref{lcGiants} for the two red giants, the red-straggler star 
and two of the RGB-AGB stars.

\section{Summary}

From an 8.3--day photometric campaign on the core of 47 Tucanae using
WFPC2 on the Hubble Space Telescope, we have for the first time
searched for variability among the main sequence stars from the
cluster turnoff at $V = 17$ down to magnitudes as faint as $V = 25$. A
total of 114 variable stars were found out of the 46422 non-saturated
stars considered.  

A total of 11 detached eclipsing binaries were
detected, thus the observed frequency of these stars is
0.02\%. Additionally we detected 15 W UMa stars corresponding to an
observed frequency of 0.03\%.  If the angular momentum loss scenario
is correct, and the distribution of binaries with log period is flat,
this results in two independent estimates of 13\% and 14\% for the 
main-sequence binary
frequency in the cluster core based on the detached eclipsing binaries
and on the W UMa's respectively.

Ten low-amplitude, short-period variables were identified that we
assume to be the non-eclipsing counterparts of the W UMa stars. 

A further 65 new variable stars have been found that are most likely
BY Dra stars in binary systems. These stars are much more strongly
concentrated towards the core of the cluster than single stars and
follow the same spatial distribution as other binary stars.

We identify a distinct class of variables, the red stragglers, found
in the (V,V-I) color-magnitude diagram slightly below and extending
redwards from the subgiant branch.  Six of these stars were found in
our survey.  We propose that these may be binary systems in which the
more massive component is losing a small amount of mass to its
companion or out of the system after reaching the subgiant stage of its
evolution.

\acknowledgements
We thank  Mario Livio, Karen Pollard, Kailash Sahu, Alison Sills, 
David Soderblom and Don Vandenberg for discussion.
This work was supported in part by STScI grant GO-8267.01-97A to
the Space Telescope Science Institute and by several STScI grants
from the same proposal to co-I institutions.

%%
%%      Here are the figures
%%
%%

\clearpage

\clearpage

\clearpage

\clearpage

\clearpage

\clearpage

\clearpage

\clearpage

\clearpage

\clearpage

\clearpage

\clearpage

\clearpage

\clearpage

\clearpage

\clearpage

\clearpage

\clearpage

\clearpage

\clearpage

\clearpage

\clearpage

\clearpage

\clearpage

\clearpage

\clearpage

%%
%%
%%      Here are the tables
%%
%%

\clearpage
\begin{deluxetable}{lllllllcc}
\tabletypesize{\small}
\rotate
\tablecaption{\label{tableEB} Eclipsing binaries. Phase is the epoch of primary minimum,
HJD - 2451300}
\tablehead{
\colhead{Name} & \colhead{Period (d)} & \colhead{Phase (d)} &
\colhead{$V$} & \colhead{$U$--$V$} &
\colhead{$V$--$I$} & \colhead{$\alpha$ \tablenotemark{1}}
& \colhead{$\delta$} }
\startdata
PC1-V17  & 1.128  & 65.920 & 18.994 & 1.068  & 0.992 & 00:24:09.2475 & -72:05:04.546 \\
WF2-V01  & 10.15  & 70.87  & 20.553 & 2.760  & 1.442 & 00:24:05.8876 & -72:04:24.816 \\
WF2-V02  & 3.97   & 67.96  & 22.137 & 1.738  & 2.525 & 00:24:04.8394 & -72:04:04.308 \\
WF2-V03  & 1.340  & 65.442 & 16.194 & 0.402  & 0.550 & 00:24:15.4294 & -72:04:41.234 \\
WF3-V01  & 0.4174 & 65.703 & 19.985 & 1.582  & 1.139 & 00:24:19.1672 & -72:04:31.339 \\
WF3-V02  & 2.31   & 67.22  & 21.297 & 2.734  & 1.672 & 00:24:30.4310 & -72:04:46.812 \\
WF4-V01  & 8.87   & 63.53  & 19.276 & 1.503  & 1.043 & 00:24:04.7093 & -72:06:08.576 \\
WF4-V02  & 4.19   & 65.62  & 20.384 & 1.663  & 1.281 & 00:24:11.8666 & -72:05:07.737 \\
WF4-V03  & 0.5875 & 65.682 & 19.147 & 0.899  & 0.931 & 00:24:18.4695 & -72:06:02.872 \\
WF4-V04  & 4.92   & 64.85  & 17.129 & 0.669  & 0.778 & 00:24:08.2956 & -72:05:35.004 \\
\enddata
\tablenotetext{1}{Coordinates in this and subsequent tables were calculated
          using the STSDAS task METRIC and archival image u5jm070cr.}
\end{deluxetable}

\clearpage
\begin{deluxetable}{llllllccc}
\tabletypesize{\small}
\rotate
\tablecaption{\label{tableWUMa} W UMa binaries}
\tablehead{
\colhead{Name} & \colhead{Period (d)} & \colhead{$V$} & \colhead{$U$--$V$} &
\colhead{$V$--$I$} & \colhead{$\alpha$} & \colhead{$\delta$} & \colhead{classification}}
\startdata
PC1-V04   & 0.3301 &  17.523 & 0.479 & 0.668 &  00:24:07.5247 & -72:04:58.241 & contact \\
PC1-V07   & 0.4188 &  18.857 & 1.012 & 0.886 &  00:24:04.9480 & -72:04:51.168 & semi-detached \\
PC1-V08   & 0.5305 &  18.383 & 0.812 & 0.842 &  00:24:03.1840 & -72:05:05.123 & semi-detached \\
PC1-V10   & 0.4309 &  16.761 & 0.324 & 0.502 &  00:24:07.6423 & -72:05:01.828 & semi-detached \\
PC1-V18   & 0.2956 &  17.596 & 0.725 & 0.770 &  00:24:07.0787 & -72:05:04.067 & contact \\
PC1-V19   & 0.4145 &  20.137 & 1.819 & 1.199 &  00:24:04.0533 & -72:05:01.221 & semi-detached \\
PC1-V20   & 0.2114 &  22.805 & 1.808 & 1.907 &  00:24:06.3777 & -72:04:49.255 & contact \\
WF2-V04   & 0.2234 &  19.250 & 1.304 & 1.014 &  00:24:09.5679 & -72:03:47.627 & contact \\
WF2-V05   & 0.2021 &  20.184 & 1.959 & 1.212 &  00:24:10.6619 & -72:03:52.601 & contact \\
WF2-V06   & 0.3747 &  20.001 & 0.892 & 0.905 &  00:24:06.0725 & -72:04:16.835 & semi-detached \\
WF2-V07   & 0.2907 &  16.654 & 1.051 & 0.841 &  00:24:09.7762 & -72:04:51.612 & semi-detached \\
WF3-V03   & 0.3311 &  17.532 & 0.403 & 0.683 &  00:24:15.0189 & -72:04:56.576 & contact \\
WF3-V04   & 0.2516 &  18.852 & 0.914 & 0.924 &  00:24:18.3647 & -72:04:55.203 & contact \\
WF3-V05   & 0.3450 &  17.040 & 0.386 & 0.626 &  00:24:24.9261 & -72:05:07.236 & contact \\
WF4-V05   & 0.2588 &  18.643 & 0.122 & 0.131 &  00:24:14.6191 & -72:05:40.371 & semi-detached \\
\enddata
\end{deluxetable}

\clearpage
\begin{deluxetable}{llllllccc}
\tabletypesize{\small}
\rotate
\tablecaption{\label{tableShort} Non-eclipsing contact and semi-detached binaries}
\tablehead{
\colhead{Name} & \colhead{Period (d)} & \colhead{$V$} & \colhead{$U$--$V$} & 
\colhead{$V$--$I$} & \colhead{$\alpha$} & \colhead{$\delta$} & \colhead{classification}}
\startdata
PC1-V21  & 0.3916 & 21.143 &  2.414 &  1.387 & 00:24:05.3106 & -72:04:56.160 & semi-detached \\
PC1-V36  & 0.3972 & 17.653 &  0.247 &  0.111 & 00:24:06.2184 & -72:04:52.168 & semi-detached \\
PC1-V46  & 0.2656 & 17.785 &  0.589 &  0.651 & 00:24:05.7776 & -72:04:49.139 & contact \\
WF2-V08  & 0.4011 & 21.277 & -1.875 & -0.280 & 00:24:15.1166 & -72:03:36.569 & semi-detached \\
WF2-V09  & 0.2405 & 20.137 & -1.246 &  0.396 & 00:24:16.7173 & -72:04:27.124 & semi-detached \\
WF3-V06  & 0.2316 & 22.929 &  0.500 &  0.697 & 00:24:25.4577 & -72:04:38.204 & semi-detached \\
WF3-V07  & 0.3163 & 19.597 & -0.066 & -0.060 & 00:24:14.2860 & -72:04:54.745 & semi-detached \\
WF3-V08  & 0.2108 & 20.832 &  2.853 &  1.202 & 00:24:15.7319 & -72:04:51.695 & semi-detached \\
WF3-V09  & 0.3311 & 19.022 &  0.723 &  0.782 & 00:24:11.9652 & -72:05:00.108 & semi-detached \\
WF3-V10  & 0.3325 & 17.120 &  0.639 &  0.732 & 00:24:13.3209 & -72:05:06.947 & contact \\
\enddata
\end{deluxetable}

\clearpage
\begin{deluxetable}{lllllllccc}
\tabletypesize{\small}
\rotate
\tablecaption{\label{tableBYDra} BY Draconis variables. $A_{V}$ is the $V$-band semi-amplitude (mag).}
\tablehead{
\colhead{Name} & \colhead{Period (d)} & \colhead{$A_{V}$} & \colhead{$V$} & \colhead{$U$--$V$} & 
\colhead{$V$--$I$} & \colhead{$\alpha$} & \colhead{$\delta$} }
\startdata
PC1-V22  & 2.50  &  0.012 & 17.839 & 0.633 & 0.724 & 00:24:04.7693 & -72:04:56.028  \\
PC1-V23  & 4.36  &  0.030 & 16.985 & 0.593 & 0.791 & 00:24:06.0691 & -72:04:52.698  \\
PC1-V24  & 2.43  &  0.024 & 17.363 & 0.589 & 0.764 & 00:24:06.9408 & -72:04:57.384  \\
PC1-V25  & 1.03  &  0.022 & 20.849 & 1.891 & 1.139 & 00:24:06.5582 & -72:04:55.926  \\
PC1-V26  & 2.43  &  0.005 & 17.653 & 0.462 & 0.713 & 00:24:06.4777 & -72:04:46.016  \\
PC1-V27  & 4.03  &  0.009 & 18.121 & 0.552 & 0.718 & 00:24:08.4830 & -72:05:00.408  \\
PC1-V28  & 1.16  &  0.036 & 20.135 & 1.351 & 1.017 & 00:24:04.6105 & -72:04:59.546  \\
PC1-V29  & 0.528 &  0.038 & 21.136 & 2.206 & 1.402 & 00:24:07.8785 & -72:05:13.195  \\
PC1-V30  & 0.798 &  0.015 & 19.279 & 1.079 & 1.073 & 00:24:03.8530 & -72:05:12.300  \\
PC1-V31  & 4.35  &  0.024 & 19.334 & 1.013 & 0.947 & 00:24:06.8886 & -72:04:48.795  \\
PC1-V32  & 1.64  &  0.008 & 18.858 & 0.910 & 0.792 & 00:24:05.4040 & -72:04:52.316  \\
PC1-V33  & 1.74  &  0.011 & 19.075 & 0.880 & 0.952 & 00:24:06.6562 & -72:04:44.424  \\
PC1-V34  & 1.38  &  0.017 & 20.382 & 1.744 & 1.118 & 00:24:06.9072 & -72:04:52.677  \\
PC1-V35  & 4.86  &  0.004 & 17.695 & 0.498 & 0.779 & 00:24:06.7297 & -72:04:42.026  \\
PC1-V37  & 4.86  &  0.010 & 19.283 & 1.255 & 1.045 & 00:24:04.2535 & -72:04:56.922  \\
PC1-V38  & 6.89  &  0.007 & 19.285 & 1.271 & 1.008 & 00:24:05.9036 & -72:04:57.097  \\
PC1-V39  & 4.73  &  0.005 & 18.435 & 0.725 & 0.757 & 00:24:04.9713 & -72:05:00.590  \\
PC1-V40  & 4.47  &  0.005 & 18.569 & 0.944 & 0.887 & 00:24:05.6368 & -72:04:53.466  \\
PC1-V41  & 5.02  &  0.002 & 17.625 & 0.867 & 0.704 & 00:24:08.4775 & -72:05:07.420  \\
PC1-V42  & 0.521 &  0.015 & 21.170 & 1.634 & 1.335 & 00:24:01.7154 & -72:05:01.649  \\
PC1-V43  & 5.16  &  0.003 & 18.933 & 0.839 & 0.924 & 00:24:06.5748 & -72:04:44.095  \\
WF2-V10  & 9.18  &  0.009 & 17.487 & 0.553 & 0.776 & 00:24:15.5759 & -72:04:33.326  \\
WF2-V11  & 0.843 &  0.074 & 20.025 & 1.299 & 1.044 & 00:24:13.7868 & -72:04:02.830  \\
WF2-V12  & 1.18  &  0.061 & 19.437 & 1.059 & 1.000 & 00:24:13.2504 & -72:04:47.190  \\
WF2-V13  & 5.93  &  0.005 & 18.066 & 0.444 & 0.713 & 00:24:13.2992 & -72:04:36.197  \\
WF2-V14  & 4.14  &  0.004 & 17.986 & 0.476 & 0.714 & 00:24:08.1418 & -72:04:26.200  \\
WF2-V15  & 1.29  &  0.007 & 18.192 & 0.642 & 0.700 & 00:24:10.8166 & -72:04:43.935  \\
WF2-V16  & 1.74  &  0.037 & 21.491 & 2.472 & 1.660 & 00:24:14.9671 & -72:03:32.018  \\
WF2-V17  & 0.633 &  0.003 & 17.338 & 0.463 & 0.753 & 00:24:06.2042 & -72:04:30.219  \\
WF2-V18  & 3.06  &  0.014 & 18.312 & 0.872 & 0.936 & 00:24:12.3851 & -72:04:22.230  \\
WF2-V19  & 1.56  &  0.033 & 21.188 & 2.083 & 1.682 & 00:24:07.0542 & -72:04:14.252  \\
WF2-V20  & 2.58  &  0.049 & 20.886 & 2.162 & 1.391 & 00:24:04.5152 & -72:03:56.537  \\
WF2-V21  & 1.50  &  0.023 & 20.800 & 2.184 & 1.428 & 00:24:09.7754 & -72:04:26.184  \\
WF2-V22  & 0.549 &  0.054 & 20.713 & 1.325 & 1.603 & 00:24:06.8901 & -72:04:20.155  \\
WF2-V23  & 7.20  &  0.008 & 18.894 & 1.313 & 0.871 & 00:24:12.2897 & -72:04:11.555  \\
WF2-V24  & 6.39  &  0.006 & 17.966 & 0.425 & 0.659 & 00:24:15.1099 & -72:04:23.037  \\
WF2-V25  & 6.62  &  0.004 & 17.712 & 0.400 & 0.678 & 00:24:04.8793 & -72:04:24.885  \\
WF2-V26  & 5.70  &  0.007 & 20.312 & 1.091 & 1.107 & 00:24:22.4470 & -72:04:05.321  \\
WF2-V27  & 2.47  &  0.003 & 18.024 & 0.636 & 0.793 & 00:24:05.7002 & -72:04:18.673  \\
WF2-V28  & 6.39  &  0.004 & 19.265 & 0.954 & 0.889 & 00:24:14.5934 & -72:04:05.350  \\
WF2-V29  & 3.76  &  0.002 & 17.547 & 0.405 & 0.700 & 00:24:11.1291 & -72:04:40.100  \\
WF3-V11  & 7.20  &  0.003 & 17.468 & 0.624 & 0.715 & 00:24:14.4423 & -72:05:05.259  \\
WF3-V12  & 1.78  &  0.006 & 17.479 & 0.493 & 0.722 & 00:24:16.4839 & -72:04:47.005  \\
WF3-V13  & 1.99  &  0.003 & 17.582 & 0.448 & 0.700 & 00:24:22.8584 & -72:05:31.781  \\
WF3-V14  & 5.16  &  0.028 & 19.669 & 1.262 & 1.033 & 00:24:21.0290 & -72:05:02.810  \\
WF3-V15  & 0.437 &  0.023 & 20.294 & 1.917 & 1.415 & 00:24:24.3744 & -72:05:00.255  \\
WF3-V16  & 2.50  &  0.010 & 19.257 & 1.066 & 0.954 & 00:24:14.7970 & -72:05:02.997  \\
WF3-V17  & 7.89  &  0.011 & 19.875 & 1.410 & 0.953 & 00:24:21.7737 & -72:04:25.483  \\
WF3-V18  & 5.70  &  0.008 & 19.728 & 1.388 & 0.921 & 00:24:26.8360 & -72:05:25.692  \\
WF3-V19  & 1.88  &  0.041 & 22.996 & 0.254 & 1.975 & 00:24:28.6891 & -72:05:00.584  \\
WF3-V20  & 7.57  &  0.013 & 21.514 & 2.808 & 1.585 & 00:24:19.8193 & -72:05:25.453  \\
WF3-V21  & 4.03  &  0.015 & 21.420 & 2.867 & 1.998 & 00:24:18.5850 & -72:05:04.426  \\
WF3-V22  & 0.893 &  0.010 & 20.764 & 3.050 & 1.501 & 00:24:15.0728 & -72:04:49.656  \\
WF3-V23  & 9.75  &  0.001 & 17.501 & 0.474 & 0.670 & 00:24:16.0377 & -72:04:59.281  \\
WF3-V24  & 7.89  &  0.020 & 22.160 &       & 1.714 & 00:24:24.5505 & -72:04:53.861  \\
WF4-V06  & 6.13  &  0.031 & 17.120 & 0.505 & 0.735 & 00:24:10.2673 & -72:05:06.501  \\
WF4-V07  & 1.44  &  0.026 & 19.894 & 1.173 & 1.138 & 00:24:12.2932 & -72:05:21.804  \\
WF4-V08  & 2.04  &  0.013 & 20.670 & 2.291 & 1.462 & 00:24:07.2830 & -72:06:19.875  \\
WF4-V09  & 1.37  &  0.019 & 20.743 & 2.111 & 1.270 & 00:24:07.4045 & -72:05:53.703  \\
WF4-V10  & 2.62  &  0.007 & 19.767 & 2.320 & 1.324 & 00:24:06.3807 & -72:05:42.025  \\
WF4-V11  & 8.72  &  0.016 & 20.604 & 1.865 & 1.142 & 00:24:14.9759 & -72:05:29.017  \\
WF4-V12  & 6.13  &  0.004 & 17.624 & 0.461 & 0.650 & 00:24:04.5026 & -72:05:37.311  \\
WF4-V13  & 9.18  &  0.013 & 20.911 &       & 1.297 & 00:24:13.0519 & -72:06:04.108  \\
WF4-V14  & 4.35  &  0.004 & 18.180 & 0.640 & 0.730 & 00:24:09.5175 & -72:06:00.822  \\
WF4-V15  & 0.903 &  0.028 & 21.813 &       & 1.734 & 00:24:08.1056 & -72:06:21.536  \\
\enddata
\end{deluxetable}

\clearpage
\begin{deluxetable}{lllllllccc}
\tabletypesize{\small}
\rotate
\tablecaption{\label{tableRS} Red Straggler variables}
\tablehead{
\colhead{Name} & \colhead{Period (d)} & \colhead{$A_{V}$} & \colhead{$V$} & 
\colhead{$U$--$V$} & 
\colhead{$V$--$I$} & \colhead{$\alpha$} & \colhead{$\delta$} }
\startdata
PC1-V11  & 1.108 & 0.117 & 17.391 & 0.287 & 0.941 & 00:24:04.6452 & -72:04:55.300  \\
PC1-V48  & 6.71 & 0.005 & 17.386 & 1.147 & 0.989 & 00:24:10.1782 & -72:04:59.780  \\
WF2-V31  & 5.34 & 0.073 & 17.576 & 1.301 & 1.139 & 00:24:14.9180 & -72:04:43.454  \\
WF2-V32  & 9.2  & 0.011 & 17.141 & 1.013 & 0.997 & 00:24:13.5415 & -72:03:34.452  \\
WF4-V17  & 4.86 & 0.003 & 17.567 & 2.573 & 1.294 & 00:24:11.5517 & -72:06:31.694  \\
WF4-V18  & 5.90 & 0.020 & 17.277 & 0.769 & 0.870 & 00:24:11.7456 & -72:05:12.414  \\
\enddata
\end{deluxetable}

\clearpage
\begin{deluxetable}{llllllccc}
\tabletypesize{\small}
\rotate
\tablecaption{\label{tableOther} Other variables}
\tablehead{
\colhead{Name} & \colhead{Period (d)} & \colhead{$V$} & \colhead{$U$--$V$} & 
\colhead{$V$--$I$} & \colhead{$\alpha$} & \colhead{$\delta$} }
\startdata
PC1-V12  &         & 16.076 & 0.318 & 0.397 & 00:24:06.1630 & -72:04:45.264 \\
PC1-V47  & 6.39    & 19.772 &-0.904 & 0.456 & 00:24:03.9773 & -72:04:57.885 \\
PC1-V52  &         & 18.514 & 0.561 & 0.615 & 00:24:05.9771 & -72:04:45.979 \\
PC1-V53  &         & 20.411 &-1.028 & 1.138 & 00:24:05.7208 & -72:04:56.015 \\
WF2-V30  & 1.16    & 21.295 &-1.624 & 0.400 & 00:24:14.8416 & -72:03:44.793 \\
WF2-V47  &         & 17.750 & 1.247 & 0.725 & 00:24:08.2107 & -72:04:27.204 \\
WF2-V48  &         & 21.677 &-0.434 & 2.657 & 00:24:15.6337 & -72:04:36.312 \\
WF4-V16  & 9.75    & 24.070 &       & 0.918 & 00:24:04.6068 & -72:06:05.958 \\
WF4-V26  &         & 17.694 & 0.571 & 0.702 & 00:24:14.5067 & -72:05:59.154 \\
\enddata
\end{deluxetable}

\clearpage
\begin{deluxetable}{llllllccc}
\tabletypesize{\small}
\rotate
\tablecaption{\label{tableRGB} Variable giants}
\tablehead{
\colhead{Name} & \colhead{Period (d)} & \colhead{$V$} & \colhead{$U$--$V$} & 
\colhead{$V$--$I$} & \colhead{$\alpha$} & \colhead{$\delta$} }
\startdata
WF4-V19   & 5.9    & 14.465 & 1.865 & 1.246 & 00:24:07.5500 & -72:05:20.097  \\  
WF4-V20   & $>$10  & 15.219 & 1.377 & 0.968 & 00:24:10.9840 & -72:05:08.778  \\  
\enddata
\end{deluxetable}

\clearpage
\begin{deluxetable}{llllllccc}
\tabletypesize{\small}
\rotate
\tablecaption{\label{tableAGB} Variable RGB-AGB stars}
\tablehead{
\colhead{Name} & \colhead{Period (d)} & \colhead{$V$} & \colhead{$U$--$V$} & 
\colhead{$V$--$I$} & \colhead{$\alpha$} & \colhead{$\delta$}}
\startdata
PC1-V49  & 7.51   & 12.152 & 3.435 & 1.440 & 00:24:08.4012 & -72:05:11.301  \\  
PC1-V50  & 7.2    & 12.108 & 3.857 & 1.364 & 00:24:10.5132 & -72:05:02.312  \\  
%PC1-S127  & 1.84   & 16.216 & 1.319 & 0.906 & 00:24:04.6238 & -72:04:46.099 & Detached eclipsing \\  
%WF2-S1051 & 0.2907 & 16.654 & 1.051 & 0.841 & 00:24:09.7762 & -72:04:51.612 & W UMa     	 \\  
WF2-V33  & 6.36   & 12.341 & 2.930 & 0.730 & 00:24:03.4806 & -72:04:17.984 	 \\  
WF2-V34  & 5.5	  & 12.266 & 3.341 & 1.203 & 00:24:08.1873 & -72:04:35.784 	 \\  
WF2-V35  & 5.7    & 12.238 & 3.025 &	    & 00:24:14.4012 & -72:04:27.294 	 \\  
WF2-V36  & 2.85   & 12.789 & 2.841 &	    & 00:24:01.6237 & -72:04:04.320 	 \\  
WF2-V37  & 6.5    & 12.245 & 3.437 &	    & 00:24:13.0092 & -72:03:55.552 	 \\  
WF2-V38  & 9      & 11.214 & 3.640 &	    & 00:24:14.5151 & -72:04:44.337 	 \\  
WF2-V39  & 6.1    & 12.432 & 3.230 &	    & 00:24:12.5795 & -72:03:38.357 	 \\  
WF2-V40  & 9      & 12.011 & 3.451 &	    & 00:24:01.6279 & -72:04:09.160 	 \\  
WF2-V41  & 6.6    & 12.195 & 3.313 &	    & 00:24:07.5314 & -72:04:23.170 	 \\  
WF2-V42  & $>$10  & 11.868 & 3.642 &	    & 00:24:11.3878 & -72:03:25.178 	 \\  
WF2-V43  & 5.9    & 12.558 & 2.748 &	    & 00:24:10.3531 & -72:04:40.807 	 \\  
WF2-V44  & $>$10  & 12.175 & 3.287 &	    & 00:24:17.0016 & -72:04:30.134 	 \\  
WF2-V45  & 5      & 12.264 & 3.100 &	    & 00:24:06.5038 & -72:04:25.938 	 \\  
WF2-V46  & 3.5    & 12.648 & 3.008 &	    & 00:24:13.5246 & -72:04:28.373 	 \\  
WF3-V25  & 6.1    & 12.132 & 3.278 & 1.400 & 00:24:28.5997 & -72:04:42.486 	 \\  
WF3-V26  & 9      & 11.919 & 3.449 & 1.534 & 00:24:13.4766 & -72:04:52.361 	 \\  
WF3-V27  & 7.5    & 11.962 & 3.579 & 1.484 & 00:24:19.0539 & -72:05:18.956 	 \\  
WF4-V21  & 7      & 12.383 & 3.016 & 1.699 & 00:24:09.9840 & -72:05:33.961 	 \\  
WF4-V22  & 5      & 12.477 & 3.046 & 1.585 & 00:24:20.0184 & -72:05:51.770 	 \\  
WF4-V23  & 3.9    & 12.530 & 3.018 & 1.353 & 00:24:06.6003 & -72:06:15.451 	 \\  
WF4-V24  & 9      & 12.741 & 2.944 & 1.271 & 00:24:11.0310 & -72:05:35.552 	 \\  
WF4-V25  & $>$10  & 12.387 & 3.126 & 1.605 & 00:24:20.2301 & -72:05:58.480 	 \\  
\enddata
\end{deluxetable}

\end{document}